\documentclass[10pt]{article}
\RequirePackage{amsthm,amsmath, bm}
\allowdisplaybreaks[4]
\usepackage{graphicx,psfrag,epsf}
\usepackage{enumerate}

\usepackage[colorlinks,
linkcolor=blue,
anchorcolor=blue,
citecolor=blue]{hyperref}
\usepackage{subfigure}
\usepackage{amsbsy, amssymb, color,arydshln}

\usepackage{mathrsfs}

\usepackage{natbib}

\usepackage{tikz-cd}
\usepackage{caption}
\usepackage{float}

\usepackage{bookmark}  

\theoremstyle{definition}
\newtheorem{assumption}{Assumption}
\newtheorem{theorem}{Theorem}
\newtheorem{lemma}{Lemma}
\newtheorem{example}{Example}
\newtheorem{definition}{Definition}
\newtheorem{condition}{Condition}

\newtheorem{proposition}{Proposition}

\begin{document}

	\def\spacingset#1{\renewcommand{\baselinestretch}
		{#1}\small\normalsize} \spacingset{1}

	
	\spacingset{1.45}

	\title{\bf Causal Inference with Unmeasured Confounding from Nonignorable Missing Outcomes}
	\author{Renzhong Zheng\\[1em]
	\normalsize  School of Statistics, Beijing Normal University, Beijing, China \\} 
    \date{May, 2023} 
	
	\renewcommand{\thefootnote}{}
	\footnotetext{Email: 202121011028@mail.bnu.edu.cn}

	\maketitle

	\bigskip
	\begin{abstract}
		Observational studies are the primary source of data for causal inference, but it is challenging when existing unmeasured confounding. Missing data problems are also common in observational studies. How to obtain the causal effects from the nonignorable missing data with unmeasured confounding is a challenge.  In this paper, we consider that how to obtain  complier average causal effect with unmeasured confounding from the nonignorable missing outcomes.   We propose an auxiliary variable which plays two roles simultaneously, the one is  the shadow variable for identification  and the other is the instrumental variable for inference. We also illustrate some difference between some missing outcomes mechanisms in the previous work and the shadow variable assumption. We give a causal diagram to illustrate this description.
		Under such a setting, we present a general condition for nonparametric identification of the full data law from the nonignorable missing outcomes with this auxiliary variable.  For inference, firstly, we recover the mean value of the outcome based on the generalized method of moments. Secondly, we propose an estimator to adjust for the unmeasured confounding to obtain  complier average causal effect. We also establish the asymptotic results of the estimated parameters. We evaluate its performance via simulations and apply it to a real-life dataset about a political analysis.        
	\end{abstract}
	
	\noindent%

	\newpage 
	\section{Introduction}
	Causal inference plays an important role in many fields, such as epidemiology, economics and social sciences. Observational studies are the primary source of data for causal inference. 
	In observational studies, if all the confounders are measured, one can use some standard methods to adjust for the confounding bias, such as stratification and matching \citep{rubin1973matching,stuart2010matching}, propensity score \citep{rosenbaum1983central}, inverse probability weighting \citep{horvitz1952generalization}, outcome regression-based estimation \citep{rubin1979using}, doubly robust estimation \citep{bang2005doubly}.  However, unmeasured confounding is present in most observational studies. In this case, one cannot identify the causal effects from the observational data without additional assumptions. Auxiliary variables are often used to identify causal effects in adjustment for unmeasured confoundering. The instrumental variable (IV) is an
	auxiliary variable to adjust for the unmeasured confounding. There are two frameworks used in the analysis of IV, one is a structural equation model \citep{wright1928tariff,goldberger1972structural}; the other one is the monotonicity assumption, which means a monotone effect of the IV on the treatment. Under the monotonicity assumption, IV can identify and estimate the compliers average causal effect  \citep{imbens1994identification,angrist1995two,angrist1996identification}.
	More details about the two framework of the instrument variable can be found in \cite{wang2018bounded}.
	
	Missing data is also a common problem in observational studies. According to \cite{rubin1976inference}, the missing mechanism is called missing at random (MAR) or ignorable if it is independent of the missing values conditional on observed data, which means the missing mechanism only depends on the observed data. Otherwise, it is called missing not at random (MNAR) or nonignorable \citep{littlerubin2002}. 
     Compared to MAR, MNAR is much more challenging. Identification is generally not available under MNAR without additional assumptions \citep{robins1997toward}. 
	In the previous papers, authors would make sufficient parametric assumptions about the full data law to ensure the validity of the identification \citep{wu1988estimation,littlerubin2002,roy2003modeling}. 
	However, as noted by \cite{miao2016identifiability} and \cite{wang2014instrumental}, the identification of fully parametric models can even fail under MNAR. 

	A new framework for identification and semiparametric inference was recently proposed by \cite{miao2016varieties}, \cite{miao2018identification}, and \cite{miao2015identification}, building on earlier work by \cite{d2010new}, \cite{kott2014calibration}, \cite{wang2014instrumental}, and \cite{zhao2015semiparametric} which studied identification of several parametric and semiparametric models. 
    A shadow variable is associated with the outcome prone to missingness but independent of the missing mechanism conditional on treatment and possibly unobserved outcome.
	In the context of missing covariate data,  \cite{miao2018identification} studied identification of generalized linear models and some semiparametric models, and then proposed an inverse probability weighted estimator that  incorporates the shadow variable to guarantee the unbiased estimation. 
	In the context of missing outcomes data, \cite{miao2016varieties} considered the identification of a location-scale model and described a doubly robust estimator.  
	If a shadow variable is fully observed, we can use it to recover the distribution of unobserved outcome	which is missing not at random with the fully observed covariates.  \cite{miao2015identification} used a valid shadow variable to nonparametric identification of the full data distribution under nonignorable missingness, and they developed the semiparametric theory for some semiparametric estimators with a shadow variable under MNAR.

	 It is highly necessary to combine causal inference with missing data research \citep{ding2018causal}. In the case of covariates missing not at random,  \cite{ding2014identifiability} showed the identification of the causal effects for four interpretable missing data mechanisms and proposed the upper and lower bound for causal effects.
	When the unmeasured confounders are missing not at random,  \cite{yang2019causal} generalized the results in \cite{ding2014identifiability} to establish a general condition for identification of the causal effects, they further developed parametric and nonparametric inference for the causal effects. 
	With the nonignorable missing outcomes, the identification and estimation of the causal effects is more challenging, different types of the missing mechanism are imposed to guarantee the identification and estimation under the principal stratification framework. \cite{frangakis1999addressing} proposed the latent ignorable (LI) missing data mechanism to guarantee the identification and estimated the complier average causal effects (CACE) . Under the LI missing data mechanism, \cite{o2005likelihood} proposed two methods based on the moments and likelihood to estimate  CACE for normally distributed outcomes. \cite{zhou2006itt} proposed both moment and maximum likelihood estimators of  CACE for the binary outcomes.
	Otherwise, some authors proposed another missing data mechanisms, which is called the
	outcome-dependent nonignorable (ODN) missing data mechanism. \cite{chen2009identifiability} and \cite{imai2009statistical}  established the identification of  CACE for discrete outcomes based on the likelihood method under the	outcome-dependent nonignorable  missing data mechanism. \cite{chen2015semiparametric} proposed an exponential family assumption about the conditional density of the outcome variable to establish semiparametric inference of  CACE for continuous outcomes under ODN missing data mechanism. \cite{li2017identifiability} established the identification with a shadow variable and estimation of causal mediation effects from the outcomes MNAR.

	 In this paper, we obtain CACE with unmeasured confounding from nonignorable missing outcomes. Different from the previous work, we impose no assumption about the missing outcomes mechanisms. Instead, we  impose a shadow variable assumption on the instrumental variable to guarantee the identification and estimate CACE. From Figure \ref{figure 3}, we allow an arrow from treatment to missing mechanism, which is more reasonable in some application cases. 
	 In Section 2, we illustrate some basic assumptions in causal inference. We also illustrate the difference between some missing data mechanisms and the shadow variable assumption.  Then we demonstrate the framework of this paper with an example.
	In Section 3, we use a shadow variable to identify the full data distribution nonparametrically under certain completeness condition. In Section 4, we apply the generalized method of moments to estimate the missing mechanism. We also establish some asymptotic results of the estimated parameters in the missing mechanism.
	Then we recover the mean value of the outcome and propose an estimator of  CACE. In Section 5, we conduct some simulation studies to evaluate the performance of some parameters estimators in missing mechanism and  the estimator of  CACE. In Section 6, we use a real data about to illustrate our approach.  In Section 7, we conclude with some discussions and provide all proofs in the Supplementary Materials.

	\section{Notation and Assumptions}
	\subsection{Potential outcomes, causal effects, basic assumptions and instrumental variable }
	
	In this paper, we consider the situation where  $\left\{\left( y_i, r_i, a_i , z_i  \right): i=1, \ldots, n\right\}$ is an independent and identically distributed sample from $\left(Y, R, A, Z\right)$.   Vectors are assumed to be column vectors, unless explicitly transposed.

	$Y$ is the outcome of interest subject to missingness. We consider the outcome $Y$ a binary variable. We let $R$ denote the missing indicator of $Y$: $R=1$ if Y is observed and $R=0$ otherwise. The observed data include $\left(A, Z\right)$ for all samples and $Y$ only for those $R=1$. We use lower-case letters for realized values of the corresponding variables, for example, $y$ for a value of the outcome variable $Y$. We use $f$ to denote a probability density or mass function. Suppose the observed data are $n$ independent and identically distributed samples.

	We use potential outcome model to define causal effects \citep{neyman1923application,rubin1974estimating}. For each unit $i$, We consider $Z$ is a binary instrument variable, where $Z_i=1$ indicates the unit $i$ is assigned to the treatment group and $Z_i=0$ indicates the unit $i$ is assigned to the control group. Let $A_i(z)$ denote the potential outcome of the unit $i$ that the instrument variable $Z$ was set to level $z$.
	 $A_i(z)=1$ indicates that unit $i$ would receive the  treatment if assigned $z$, and $A_i(z)=0$ indicates that unit $i$ would receive the control if assigned $z$.  Similar to the definition of $A_i(z)$, we let $Y_i(z, A_i(z))$  denote the potential outcome for an unit $i$  if exposed to treatment $A_i(z)$ after $Z$ was set to level $z$. For simplicity, we let $Y_i(z)$ and $R_i(z)$ denote the potential outcomes after $Z$ was set to level $z$, respectively.   Let $Z_i$, $A_i$, $Y_i$, $R_i$  denote their observations for $i=1,\dots,n$.  
	
	The following assumption is standard in causal inference with observational studies \citep{rubin1980randomization,angrist1996identification}.
	\begin{assumption} \label{assumption 1}
		Stable Unit treatment value Assumption (SUTVA): There is no  interference between units, which means a unit’s potential outcomes cannot be affected by the treatment status of other units and  there is only one version of potential outcome of a certain treatment.
	\end{assumption}
	Under  SUTVA assumption, the definition of the potential outcome is reasonable. But when $Z_i \neq A_i$, there exists noncompliance. Under the principal stratification framework \citep{angrist1996identification,frangakis2002principal}, we define $U_i$ as the compliance status variable of unit $i$:
$$
U_i= \begin{cases} at, & \text { if } A_i(1)=1 \text { and } A_i(0)=1; \\ cp, & \text { if } A_i(1)=1 \text { and } A_i(0)=0; \\ df, & \text { if } A_i(1)=0 \text { and } A_i(0)=1; \\ nt, & \text { if } A_i(1)=0 \text { and } A_i(0)=0. \end{cases}
$$
Because $A_i(1)$ and $A_i(0)$ each can take two values, the compliance status variable $U_i$ has four different values, $nt$ for never takers, $at$ for always takers, $cp$ for compliers,
and $df$ for defiers. Because we can not observe $A_i(1)$ and $A_i(0)$ jointly, the compliance
behavior of an unit is unknown, so $U_i$ is an observed variable and it can be viewed as an unmeasured confounder. 

		\begin{definition} \label{complier average causal effect}
		The Complier Average causal effect (CACE) is defined as $CACE=E\left[Y_i(1)-Y_i(0) \mid U_i=cp \right]$.
	\end{definition}

	Because for the compliers, $Z_i=A_i$, so  CACE is a subgroup causal effect for the compliers, with incompletely observed compliance status.
	\begin{assumption}\label{assumption 2}
		Randomization:   $(Y(0), Y(1), A(0), A(1))$ is independent of $Z$.
	\end{assumption}
	 Randomization means that $Z \perp \!\!\! \perp (Y(0), Y(1), A(0), A(1)) $. 
	\begin{assumption}\label{assumption 3}
	 Monotonicity:	$A_i(1) \geq A_i(0)$ for each unit $i$.
	\end{assumption} 
	 Monotonicity assumption means there is no defiers in the population. In some studies, the monotonicity assumption is plausible when the treatment assignment has a  nonnegative effect on the treatment received for each unit. 
	 \begin{assumption} \label{assumption 4}
	 	 Nonzero average causal effect of $Z$ on $A$: The average causal effect of $Z$ on $A$, $E\left[A_i(1)-A_i(0)\right]$, is not equal to zero.
	 \end{assumption}
	\begin{assumption} \label{assumption 5}
	 Exclusion restrictions among never takers and always takers:	
		$Y_i(1)=Y_i(0)$  if $ U_i=nt$  and $Y_i(1)=Y_i(0)$ if $ U_i=at$.
	\end{assumption}
  Assumption \ref{assumption 5} means that the instrument variable $Z$ only affects the outcome through treatment and has no direct effect. 
	
	 	If the outcome $Y$ can be fully observed,  CACE can be identified and estimated by \cite{angrist1996identification}. With the nonignorable missing outcome, some assumptions about the missing data mechanisms are needed to estimate  CACE.  We will compare some assumptions about the missing data mechanisms in the next subsection.
	
	\subsection{Some Missing data mechanisms of the outcomes}
	  Before illustrating the missing data mechanisms, Assumption \ref{assumption 2} and Assumption \ref{assumption 5}  are  replaced by  Assumption \ref{assumption 6} and Assumption \ref{assumption 7} in some previous work, respectivitly. Assumption \ref{assumption 6} and Assumption \ref{assumption 7} are usually combined with the missing outcomes mechanisms in the previous work. However, we only need  Assumption \ref{assumption 2} and Assumption \ref{assumption 5} to estimate CACE in this paper, which are the weak version of Assumption \ref{assumption 6} and Assumption \ref{assumption 7}.
	
		
	
	

\begin{assumption} \label{assumption 6}
	Complete Randomization: The treatment assignment $Z$ is completely randomized.
\end{assumption}
 Complete Randomization means that $Z \perp \!\!\! \perp \{A(1), A(0), Y(1), Y(0), R(1), R(0)\} $.
	\begin{assumption} \label{assumption 7}
	Compound exclusion restrictions: For never takers and always takers, we assume that $f\{Y(1), R(1) \mid U=nt\}=f\{Y(0), R(0) \mid U=nt\}$, and $f\{Y(1), R(1) \mid U=at\}=f\{Y(0), R(0) \mid U=at\}$.
\end{assumption}
Different from traditional Assumption \ref{assumption 5}, \cite{frangakis1999addressing}  extended it to the compound exclusion restrictions. Assumption \ref{assumption 7} is stronger than Assumption \ref{assumption 5}. Under Assumption \ref{assumption 6}, Assumption \ref{assumption 7} is equivalent to $f(Y, R \mid Z=1, U=nt)=f(Y, R \mid Z=0, U=nt)$ and $f(Y, R \mid Z=1, U=at)=f(Y, R \mid Z=$ $0, U=at)$.

	 In previous work, some authors proposed the Latent  ignorability (LI) assumption \citep{frangakis1999addressing,osius2004association,zhou2006itt}: 
	 \begin{assumption}\label{assumption 8}
	 Latent  ignorability  assumption: $f\{R(z) \mid Y(z), A(z), U\}= f\{R(z) \mid U\}$ 
	 \end{assumption}
 Latent ignorability implies that for given the each principal stratum, the potential outcomes are independent of the missing indicator, which means the missing data mechanism does not depend on the missing outcome. We gave a graph model to illustrate the LI missing data mechanism.
\begin{center}
	\begin{tikzcd}[ampersand replacement=\&,row sep=large,column sep=normal]
		\&  \&                         \& U \arrow[ld,->,>=Stealth] \arrow[rd,->,>=Stealth] \arrow[dd,->,>=Stealth] \&   \\
		Z \arrow[rr,->,>=Stealth] \&  \& A \arrow[rr,->,>=Stealth] \arrow[rd,->,>=Stealth] \&                                    \& Y \\
		\&  \&                         \& R                                  \&  
	\end{tikzcd}
\end{center}
\captionof{figure}{A  graph model for the LI missing data mechanism under Assumption \ref{assumption 6} and \ref{assumption 7}} \label{figure 1}

\cite{chen2009identifiability} and \cite{chen2015semiparametric} proposed an Outcome-dependent nonignorable (ODN) missing data mechanism:
\begin{assumption}\label{assumption 9}
	Outcome-dependent nonignorable  assumption: For all $y ; z=0,1 ; a=0,1$; and $u \in\{at, cp, nt\}$, assuming
	$$
	\begin{aligned}
		P\{R(z)=1 \mid Y(z)=y, A(z)=a, U=u\} & =P\{R(z)=1 \mid Y(z)=y\} \\
		P\{R(1)=1 \mid Y(1)=y\} & =P\{R(0)=1 \mid Y(0)=y\} .
	\end{aligned}
	$$
\end{assumption}
Under the completely randomization Assumption \ref{assumption 6}, Assumption \ref{assumption 9} becomes $P(R = 1 \mid Y = y, A = a, U = u, Z = z) = P(R = 1 \mid Y = y,Z = z)$ and $P(R = 1 \mid Y = y, Z = 1) = P(R = 1 \mid Y = y,Z = 0)$. 
 	Outcome-dependent nonignorable  assumption means that $R$ depends on $Y$, but is independent of $(Z, A, U)$ given $Y$. Under the ODN missing data mechanism, the missing data indicator depends on the possibly missing outcome Y, which may be more reasonable than  LI missing data assumption in some applications. We gave a graph model to illustrate  ODN missing data mechanism.
 	\begin{center}
 		\begin{tikzcd}[ampersand replacement=\&,row sep=large,column sep=normal]
             \&  \&              \& U \arrow[ld,->,>=Stealth] \arrow[rd,->,>=Stealth] \&              \&  \&   \\
           Z \arrow[rr,->,>=Stealth] \&  \& A \arrow[rr,->,>=Stealth] \&                         \& Y \arrow[rr,->,>=Stealth] \&  \& R 
 		\end{tikzcd}
 	\end{center}
 	\captionof{figure}{A  graph model for the ODN missing data mechanism under Assumption \ref{assumption 6} and \ref{assumption 7}} \label{figure 2}

More examples and details about  LI assumption and ODN assumption can be found in \cite{chen2009identifiability} and \cite{chen2015semiparametric}.


	\subsection{The framework of this paper}
	In this paper,  we are interested in CACE with unmeasured confounding $U$ from  the nonignorable missing outcomes $Y$.

	In fact, LI assumption and ODN assumption usually need the Assumption \ref{assumption 6} and Assumption \ref{assumption 7}, which are stronger than Assumption \ref{assumption 2} and Assumption \ref{assumption 5}. In this paper, we only need the weak randomization Assumption \ref{assumption 2} and exclusion restriction  Assumption \ref{assumption 5} to estimate CACE.
	Different from the above two missing outcomes mechanisms,  we adapt the shadow variable assumption to guarantee the identification under the nonignorable missing outcomes in this paper.

	We suppose that an auxiliary variable $Z$ called a shadow variable  is fully observed  if it statisfies the following assumption \citep{miao2016varieties,miao2015identification}. 
	\begin{assumption} \label{shadow variable}
		(a): $Z \perp \!\!\! \perp R \mid(Y,  A)$;   (b): $Z \not \perp \!\!\! \perp Y \mid(R=1,  A)$.
	\end{assumption}
	
	Assumption \ref{shadow variable} implies that the shadow variable is associated with the outcome when the outcome is observed, but it is independent of the missing mechanism conditional on fully observed variables and possibly unobserved outcome \citep{d2010new,kott2014calibration,wang2014instrumental,zhao2015semiparametric}. Therefore, Assumption \ref{shadow variable} allows for the data missing not at random.

	In this paper, we suppose an auxiliary variable $Z$ satisfies both Assumptions \ref{assumption 1}-\ref{assumption 5} and Assumption \ref{shadow variable}, which plays two roles of the instrument variable and the shadow variable simultaneously. Different from the ODN missing data mechanism \ref{figure 2}, we allow the arrow from $A$ to $R$ in Figure \ref{figure 3}, which means the treatment would affect the missing indicator. It is more reasonable in some applications. We present a example to help readers to understand the framework of Fiugre \ref{figure 3}. Figure \ref{figure 3} illustrates the framework of this paper.
	
	\begin{center}
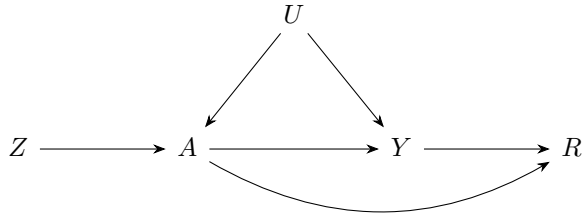

		\begin{tikzcd}[ampersand replacement=\&,row sep=normal,column sep=normal]
			\&  \&  \& U \arrow[ldd,->,>=Stealth] \arrow[rdd,->,>=Stealth] \&  \&  \&   \\
			\&  \&                                                                  \&  \&                           \&  \&   \\
			Z \arrow[rr,->,>=Stealth] \&  \& A \arrow[rrrr,->,>=Stealth,bend right=30] \arrow[rr,->,>=Stealth]                                                     \&  \& Y \arrow[rr,->,>=Stealth]              \&  \& R
		\end{tikzcd}
	\end{center}
	\captionof{figure}{A directed acyclic graph model for the auxiliary variable $Z$}\label{figure 3}

	\begin{example} \label{example 3}
     We consider a real data analysis in \cite{esterling2011estimating}. As noted in \cite{esterling2011estimating},  the authors aimed to assess that participating in an Online-chat session  whether produced higher levels of citizen efficacy among the constituents. In this paper, we choose the instrument variable $Z$ indicating  political knowledge, which means $Z$ equals to $1$ for the constituents who have high political knowledge and $0$ for low political knowledge. The treatment variable $A$ equals to $1$ for the constituents who participated in the Online-chat session and $0$ for otherwise. A week after each session, a company administered a follow-up survey to the constituents. The outcome variable $Y$ is Officials care, denoting the attitudes of the constituents about a question in the follow-up survey, which includes ``agreement'', ``disagreement'' and so on. However, some subjects may not respond the follow-up survey, which the outcomes have missing values. At the same time, it is reasonable that whether or not to participate in the session would affect the missing mechanism of the outcomes. In Section 6, we illustrate this example in more detail.
	\end{example}


	In the next section, we state that the key of identification of $f(y, r \mid a, z)$ is Assumption \ref{shadow variable}. If Assumption \ref{shadow variable} is violated, which means the shadow variable is not available, the identification of $f(y, r \mid a, z)$ is not guaranteed, even if the  parametric missingness mechanisms may not be identifiable \citep{miao2016identifiability,wang2014instrumental}. We also illustrate  extra conditions to guarantee the identification.

	\section{Identification}
	
 We aim to identify the joint distribution $f(a , y, r, z)$. We say the joint distribution $f(a , y, r, z )$ is identifiable , if and only if the joint distribution can be uniquely determined by the observed data distribution $f(y, r=1 \mid a , z ), f(r=0 \mid a, z )$ and $f(a, z )$.

	The joint distribution $f(a, y, z, r)$ can be factorized as 
	\begin{equation} \label{equation 1}
		f(a, y, z, r)= f(y, r \mid a, z) f(a, z), \tag{3.1}
	\end{equation}
	because  the variables $(A,Z)$ can be fully observeed, so $f(a,z)$ is identifiable without additional assumptions, we focus on the identification  of $f(y, r \mid a, z)$. The observed data law is captured by  $f(y, r=1 \mid a, z)$, $f(r=0 \mid a, z)$ and $f(a, z)$, which are functionals of the joint law $f(a, y, z, r)$. 
	Then we factorize $f(y, r \mid a, z)$ as
	\begin{equation} \label{equation 2}
		f(y, r \mid a, z)=f(y \mid r, a, z) f(r \mid a, z).\tag{3.2}
	\end{equation}  
	According to equation (\ref{equation 2}),  $f(y \mid r, a, z)$ represents the outcome distribution for different data patterns: $R=1$ for the observed data and $R=0$ for the missing data.
	Although $f(y \mid r=1, a, z)$ can be obtained from the observed data completely, the missing data distribution $f(y \mid r=0, a, z)$ can not be obtained directly from the observed data under MNAR.  
	The fundamental identification challenge in missing data problems is how to recover the full data distribution $f(a, y)$ and missingness process (or propensity score)$f(r \mid a, z)$ from the observed data distribution.

	We use the odds ratio function which encodes the deviation between the observed data and missing data distributions to measure the missingness process \citep{little1993pattern,little1994class}. 
	
	\begin{equation} \label{equation 3}
		\operatorname{OR}(a, y, z)=\frac{f(y \mid r=0, a, z) f(y=1 \mid r=1, a, z)}{f(y \mid r=1, a, z) f(y=1 \mid r=0, a, z)}.\tag{3.3}
	\end{equation}
	
	\vskip.1in
	Here, we use $y=1$ as a reference value, the analyst can use any other value within the support of $Y$. The odds ratio function can be applied to impose a known relationship of the outcome $Y$ \citep{little1993pattern,little1994class}.  In proposition \ref{proposition 1}, the odds ratio function plays a center role under data MNAR and it can be identified by a shadow variable.

	In the following, we hold that $\operatorname{OR}(a, y)$ and $E\left[\operatorname{OR}(a, y, z) \mid r=1, a, z\right]<+\infty$. According to the previous work, we factorize the conditional  density  function $f(y, r \mid a, z)$  as the odds ratio function and two baseline distributions       \citep{osius2004association,chen2003note,chen2004nonparametric,yun2007semiparametric,kim2011semiparametric,miao2016varieties,miao2015identification}, we establish some results  in the proposition \ref{proposition 1} based on a valid shadow variable.

	\begin{proposition} \label{proposition 1}
		Under Assumption \ref{shadow variable}, we have that for all $(A,Y,Z)$
		\begin{equation}\label{equation 4}
			\operatorname{OR}(a, y, z)=\operatorname{OR}(a, y) \equiv \frac{f(r=0 \mid a, y) f(r=1 \mid a, y=1)}{f(r=1 \mid a, y) f(r=0 \mid a, y=1) }, \tag{3.4}
		\end{equation}
		\vskip.1in
		\begin{equation}\label{equation 5}
			\begin{gathered}
				f(y, r \mid a, z)=c(a, z) f(r \mid a, y=1) f(y \mid r=1, a, z)\{\operatorname{OR}(a, y)\}^{1-r},\\[2mm]
				c(a, z)=\dfrac{f(r=1 \mid a)}{f(r=1 \mid a, y=1)} \frac{f(z \mid r=1, a)}{f(z \mid a)}, \tag{3.5}
			\end{gathered}
		\end{equation}
		\vskip.1in
		\begin{equation}\label{equation 6}
			f(r=1 \mid a, y=1)=\dfrac{E\left[\operatorname{OR}(a, y) \mid r=1, a\right]}{f(r=0 \mid a) / f(r=1 \mid a)+E\left[\operatorname{OR}(a, y) \mid r=1, a\right]}. \tag{3.6}
		\end{equation}
	\end{proposition}
	\vskip.05in
	The proof of the proposition \ref{proposition 1} is found in the Supplementary Materials. From identity (\ref{equation 4}), we demonstrate that the odds ratio function $\operatorname{OR}(a, y)$ is the function of the variables $(A,Y)$ under Assumption \ref{shadow variable}, which means the odds ratio function doesn't depend on the variable $Z$ under the Assumption \ref{shadow variable}. So we denote the odds ratio function $\operatorname{OR}(a, y, z)$ by $\operatorname{OR}(a, y)$. 
	According to equation (\ref{equation 4}), we notice that the odds ratio function also consists of the propensity score $f(r=1 \mid a, y)$, which is influenced by the  outcome itself. The odds ratio function can also be used to measure whether the selection bias exists, for instance,  the value of $\operatorname{OR}(a, y)$ represents the  deviation of missingness mechanism from the MAR \citep{miao2015identification}. 
	
	According to \cite{miao2016varieties}, we suppose throughout that $\operatorname{OR}(a, y)$ is correctly specified, which can be achieved by specifying a relatively flexible model, or following the approach suggested by \cite{higgins2008imputation} if information on the reasons for missingness are available. 
	
	
	Equation (\ref{equation 5}) plays the key role of the factorization of $f(y, r \mid a, z)$, which is factorized by the propensity score $f(r \mid a, y=1)$ evaluated at the reference level $Y = 1$, the outcome distribution $f(y \mid r=1, a, z)$ among the complete cases, and the odds ratio function $\operatorname{OR}(a, y)$. The former two is referred to as the baseline propensity score and the baseline outcome distribution, respectively. In the previous illustration, we have that $f(y \mid r=1, a, z)$ is uniquely determined from complete cases, according to equation (\ref{equation 5}) and (\ref{equation 6}), we aim to identify $f(y, r \mid a, z)$ by proposing the odds ratio function $\operatorname{OR}(a, y)$, which means the identification of the odds ratio function $\operatorname{OR}(a, y)$ is the fundamental problem in the whole framework. Propositon \ref{proposition 2} indicates some further results from identity (\ref{equation 5}) and (\ref{equation 6}) \citep{miao2015identification}.

	\begin{proposition} \label{proposition 2}
		Under Assumption \ref{shadow variable}, we have that
		\begin{equation} \label{equation 7}
			\begin{aligned}
				f(r=1 \mid a, y) & =f(r=1 \mid a, y, z), \\[2mm]
				& =\dfrac{f(r=1 \mid a, y=1)}{f(r=1 \mid a, y=1)+\operatorname{OR}(a, y) f(r=0 \mid a, y=1)}, 
			\end{aligned}
			\tag{3.7}
		\end{equation}
		\vskip.1in
		\begin{equation} \label{equation 8}
			f(y \mid r=0, a, z)=\dfrac{\operatorname{OR}(a, y) f(y \mid r=1, a, z)}{E\left[\operatorname{OR}(a, y) \mid r=1, a, z\right]}, \tag{3.8}
		\end{equation}
		\vskip.1in
		\begin{equation} \label{equation 9}
			\begin{gathered}
				E\left[\widetilde{\operatorname{OR}}(a, y) \mid r=1, a, z\right]=\frac{f(z \mid r=0, a)}{f(z \mid r=1, a)},\\[3mm]
				\text{where} \quad \widetilde{\operatorname{OR}}(a, y)=\dfrac{\operatorname{OR}(a, y)}{E\left[\operatorname{OR}(a, y) \mid r=1, a\right]}, \tag{3.9}
			\end{gathered}
		\end{equation}
		\vskip.1in
		\begin{equation} \label{equation 10}
			\operatorname{OR}(a, y)=\dfrac{\widetilde{\operatorname{OR}}(a, y)}{\widetilde{\operatorname{OR}}(a, y=1)}.  \tag{3.10}
		\end{equation}
	\end{proposition}
	\vskip.05in
	The proof of the proposition \ref{proposition 2} is found in the Supplementary Materials. The equations (\ref{equation 7}) - (\ref{equation 10}) indicate the key role of the odds ratio function in the identification framework of this paper. Equation (\ref{equation 7}) reveals that the propensity score $f(r=1 \mid a, y)$ is a function of the odds ratio function $\operatorname{OR}(a, y)$. Equation (\ref{equation 8}) shows that under the shadow variable Assumption \ref{shadow variable}, the recovery of the missing data distribution $f(y \mid r=0, a, z)$ is achievable by integrating the odds ratio function $\operatorname{OR}(a, y)$ with the complete-case distribution, therefore the full data distribution is available. Equation (\ref{equation 9}) is a Fredholm integral equation of the first kind, which $\widetilde{\operatorname{OR}}(a, y)$ is to be solved for, because $f(z \mid r=0, a)$, $f(z \mid r=1, a)$ and $f(y \mid r=1, a, z)$ can be obtained from the observed data. Equation (\ref{equation 10}) means $\operatorname{OR}(a, y)$ is related to $\widetilde{\operatorname{OR}}(a, y)$, therefore  the identification of $\operatorname{OR}(a, y)$ is equivalent to the unique solution of equation (\ref{equation 9}),
	which is guaranteed by a completeness condition of $f(y \mid r=1, a, z)$ \citep{miao2015identification}.
	
	\begin{condition} \label{condition 1}
		(The completeness  of $f(y \mid r=1, a, z)$) \qquad The function  $f(y \mid r=1, a, z)$ is complete if and only if for any square-integrable function $h(a, y)=0$ almost surely, $E\left[h(a, y) \mid r=1, a, z\right]=0$ almost surely.  
	\end{condition} 
	
	The completeness condition is also an essential condition in other identification analyses \citep{newey2003instrumental}, such as  nonparametric instrumental variable regression \citep{darolles2011nonparametric}, outcome  missing not at random \citep{miao2015identification} and confounders missing not at random \citep{yang2019causal}. In contrast to previous authors, some conditions about the shadow variable could not be justified empirically. For instance, a completeness condition is required by the full data distribution \citep{d2010new}; a monotone
	likelihood ratio is required by the full data distribution  \citep{wang2014instrumental};  a generalized linear model is considered for the full data distribution \citep{zhao2015semiparametric}. 
	As noted by \citep{miao2015identification}, the completeness condition of this paper only involves the observed data distribution $f(y \mid r=1, a, z)$, which means that it can be justified and it does not need additional model assumptions on the missing data distribution.

	In fact, Condition \ref{condition 1}  implicitly indicates that $Z$ has a larger support than $Y$. 
	However, as noted by \cite{miao2016varieties}, a binary shadow variable can not guatantee the identification of the full data distribution for a continuous outcome, which is given a counterexample. If one wants to identify a continuous outcome, a continuous shadow variable and extra conditions are needed to impose. More details about the completeness condition can be found in \cite{miao2016varieties}. The authors applied some commonly-used parametric and semiparametric models such as exponential families and location-scale families to the analysis.

	Under Assumption \ref{shadow variable}, Condition \ref{condition 1} is sufficient to ensure the unique solution from equation (\ref{equation 9}), and thus according to equation (\ref{equation 10}), the odds ratio function $\operatorname{OR}(a, y)$ is identifiable. We state the result in the following theorem.
	\begin{theorem}\label{theorem 1}
		Under Assumption \ref{shadow variable} and Condition \ref{condition 1}, the joint distribution $f(a, y, z, r)$  is identifiable.
	\end{theorem}
	The proof of the Theorem \ref{theorem 1} can be found in the Supplementary Materials. Theorem \ref{theorem 1} indicates nonparametric identification of the full data law $f(a, y, z, r)$ under MNAR is achieved with a valid shadow variable. The odds ratio function is the basis of our identification analysis. Because  $f(y \mid r=1, a, z)$ and $f(z \mid r=1, a)$ are identifiable from the observed data, thus we turn the identification of the odds ratio function $\operatorname{OR}(a, y)$ into the problem of solving for $\widetilde{\operatorname{OR}}(a, y)$ from (\ref{equation 9}). Condition \ref{condition 1} guarantees the unique solution of (\ref{equation 9}). According to equation (\ref{equation 8}), the missing data distribution $f(y \mid r=0, a, z)$ can be recovered after the odds ratio function is identifiable. And then according to equation (\ref{equation 1}) and (\ref{equation 2}), $f(y, r \mid a, z)$ and its functionals can be identified. The shadow variable plays the key role in the identification of the odds ratio function. With a valid shadow variable, nonparametric identification of the full data distribution $f(a, y, z, r)$ is achieved via the pattern-mixture factorization. Furthermore, the shadow variable is the basis of the identification of the odds ratio function, this approach guarantees equation (\ref{equation 9}) available. More details and examples about the shadow variable can be found in \cite{miao2016varieties} and \cite{miao2015identification}. 
	
	In the next section, we apply the generalized method of moments (GMM) to obtain the mean value of the outcome $Y$. Some results of the consistency and the asymptotic properties of the estimators are established. After recovering the mean value of the outcome, we propose an estimator to adjust for the unmeasured confounding to obtain the CACE.

	\section{Estimation and Inference}
	\subsection{Estimating the mean value of the nonignorable missing outcomes}
	
	We consider the situation where $Z$ is an auxiliary variable taking values of $l = 0,1$. 
	The first step of the estimation is to recover the mean value of the outcome which is missing not at random. We want to estimate the population mean $\mu = E \left[ Y \right]$ from the observed data. After the identification is guaranteed, the joint distribution of $Y$ and $R$ given $\left(A, Z\right)$  is determined by $f(y \mid a, z )$ and the missing mechanism $f(r \mid y, a, z)$. 
	\begin{equation}\label{equation 11}
		f(y, r \mid a, z)=f(y \mid a, z) f(r \mid y, a, z) . \tag{4.1}
	\end{equation}
	The missing mechanism $\pi( y, a, z) = f(r=1 \mid y, a, z) $ is the key in the process of the estimation. The conditional probability $\pi( y, a, z)$ is also called the nonresponse
	mechanism or the propensity of missing data in some literatures \citep{wang2014instrumental,shao2016semiparametric,zhang2018generalized}. For the outcome $Y$ missing not at random, the propensity depends on both observed data and missing data. Some authors imposed some parametric assumptions on both propensity and outcome model to establish some likelihood methods \citep{greenlees1982imputation,baker1988regression}.But it is sensitive for the estimators under the fully parametric models. Some authors proposed some semiparametric approaches.
	\cite{qin2002estimation} imposed a parametric model on the propensity , which is difficult to verify under nonignorable missingness. Some authors considerd some weak assumption to the model. \cite{kim2011semiparametric} proposed a semiparametric logistic regression model for the propensity . \cite{tang2003analysis} proposed a pseudo-likelihood method, imposing a parametric model for propensity but an unspecified propensity .  
	\cite{zhao2015semiparametric} consider a generalized linear model for the estimation allowing for a nonparametric missing mechanism.
	Some authors proposed an empirical likelihood method to estimate the parameters in the missing mechanism with nonignorable missing data  \citep{zhao2013empirical,tang2014empirical,niu2014empirical}.

	\cite{wang2014instrumental} applied the generalized method of moments to estimate the parameters of the missing mechanism. \cite{shao2016semiparametric} imposed an exponential tilting model on the propensity and estimated the tilting parameter and population mean
	in two steps. \cite{zhang2018generalized} proposed an approach that the parameters of interest and the tilting parameter can be estimated simultaneously with generalized method of moments and kernel regression.

	In this paper, we apply the generalized method of moments (GMM) to estimate the parameters of the missing mechanism\citep{hansen1982large,hall2004generalized}. For estimation, we consider a parametric model for the propensity $\pi( y, a, z)$ and we don't impose any parametric assumption on the outcome model $f(y \mid a, z)$, which means the $f(y \mid a, z)$ is nonparametric.

	Here we assume the propensity $\pi( y, a, z) = f(r=1 \mid y, a, z)$ satisfies the following conditions:
	\begin{condition} \label{condition 2}
		\begin{equation}
			\begin{aligned} \label{equation 4.2}
				\pi( y, a, z) &=\pi( y, a, z, \boldsymbol{\theta}) = f(r=1 \mid y, a, z, \boldsymbol{\theta}) \\[1mm] 
				&= f(r=1 \mid y, a, \boldsymbol{\theta} ) 
				=\psi\left(\alpha+\beta y+ \gamma a  \right), 
			\end{aligned}
			\tag{4.2}
		\end{equation}
		where $\boldsymbol{\theta} = \left(\alpha, \beta, \gamma \right)^{\mathrm{T}}$ is a $p$-dimensional unknown parameter. $\psi(\cdot)$ is a known, strictly monotone, and twice differentiable function from $\mathcal{R}$ to $(0,1]$.
	\end{condition}
	Similar to the discussion of \cite{wang2014instrumental} , we set a parametric propensity in which the nonresponse instrument $Z$ does not appear. However once the identification is guaranteed in Section 3, the Condition (C2) in \cite{wang2014instrumental} and \cite{zhang2018generalized} for the identiability conditions is not essential.  The next step is to estimate these parameters using the observed data.

	For applying the GMM, we need to construct a set of $q$ estimating equations:   
	\begin{equation} \label{equation 13}
		G(y,r,a, z,\boldsymbol{\theta})=\left(g_1(y, r, a,z, \boldsymbol{\theta}), \cdots, g_q(y, r, a, z, \boldsymbol{\theta})\right)^{\mathrm{T}}, \quad  \boldsymbol{\theta} \in \Theta, \tag{4.3}
	\end{equation}
	where  $\Theta$ is the parameter space containing the true parameter value $\boldsymbol{\theta}_0 = \left(\alpha_0, \beta_0, \gamma_0\right)^{\mathrm{T}}$, $\boldsymbol{\theta}$ is a $p$-dimensional vector of the parameters that we want to estimate. The estimating equations need to satisfy $E\left[g_m\left(y, r, a, z, \boldsymbol{\theta}_0)\right)\right]=0$, \quad $m=1,\cdots,q$ and $q \geq p$.  
	Let
	\begin{equation} \label{equation 14}
		\widehat{G}_{n}(\boldsymbol{\theta})=\left(\frac{1}{n} \sum\limits_{i=1}^{n} g_1(y_i, r_i, a_i ,   z_i, \boldsymbol{\theta}), \cdots, \frac{1}{n} \sum\limits_{i=1}^{n} g_q(y_i, r_i, a_i ,   z_i, \boldsymbol{\theta})\right)^{\mathrm{T}}, \quad  \boldsymbol{\theta} \in \Theta. \tag{4.4}
	\end{equation}
	$\widehat{G}_{n}(\boldsymbol{\theta})$ is the sample version of the estimating equations (\ref{equation 13}). The number of estimating equations $q$ is often greater than $p$, which implies that there is no solution to
	\begin{equation*} \label{equation 15}
		\bar{g}_m\left( \boldsymbol{\theta}\right) \equiv \frac{1}{n} \sum\limits_{i=1}^n g_m\left(y_i, r_i, a_i ,   z_i, \boldsymbol{\theta}\right)=0, \quad  m=1,\cdots,q. \tag{4.5}
	\end{equation*} 
	The best we can do is to make it as close as possible to zero by minimizing the quadratic function
	\begin{equation} \label{equation 16}
		{Q}_n( \boldsymbol{\theta})= [\widehat{G}_{n}(\boldsymbol{\theta})]^{\mathrm{T}} W [\widehat{G}_{n}(\boldsymbol{\theta})], \tag{4.6}
	\end{equation}
	where $W$ is a positive semi-definite and symmetric $q \times q$ matrix of weights. We use two-step GMM to estimate the parameters $\boldsymbol{\theta}$.

	The first step: Let $W=I_{q \times q}$ in (\ref{equation 16}), $I_{q \times q}$ is a $q \times q$ identity matrix, then we obtain $\tilde{\boldsymbol{\theta}}$ by minimizing ${Q}_n( \boldsymbol{\theta})= [\widehat{G}_{n}(\boldsymbol{\theta})]^{\mathrm{T}} [\widehat{G}_{n}(\boldsymbol{\theta})]$, which means $\tilde{\boldsymbol{\theta}} = \mathop{\arg\min}\limits_{\boldsymbol{\theta} \in \Theta} Q_n(\boldsymbol{\theta})$. 
	
	The second step: Then we obtain $\widehat{W} = W(\tilde{\boldsymbol{\theta}})$, plugging in $\widehat{W}$ in (\ref{equation 16}). Finally, we obtain the GMM estimator $\widehat{\boldsymbol{\theta}}$ by minimizing $\widehat{Q}_n( \boldsymbol{\theta})= [\widehat{G}_{n}(\boldsymbol{\theta})]^{\mathrm{T}} \widehat{W} [\widehat{G}_{n}(\boldsymbol{\theta})]$ over $\boldsymbol{\theta} \in \Theta$, which means $\widehat{\boldsymbol{\theta}}=\mathop{\arg\min}\limits_{\boldsymbol{\theta} \in \Theta} \widehat{Q}_n(\boldsymbol{\theta})$.
	
	The next step is to construct the estimating equations to estimate $\boldsymbol{\theta} = \left(\alpha, \beta, \gamma \right)^{\mathrm{T}}$. $Z$ is a discrete variable taking values of $l = 0,1$.  We let $\boldsymbol{k}$ be a $L$-dimensional column vector whose $l$-th component is $I(z=l)$ $(L=2)$, where $I(\cdot)$ is the indicator function. Similar to $\boldsymbol{k}$, we let $\boldsymbol{j}$ also be a $L$-dimensional column vector whose $l$-th component is $I(a=l)$ .
	
	We now consider the general situation. The estimating equations in (\ref{equation 13}) can be constructed by the following $q$ functions 
	
	\begin{equation} \label{equation 4.7}
		G(y, r, a, z, \boldsymbol{\theta})=\left(\begin{array}{l}
			\boldsymbol{k} \left[\dfrac{r}{ \pi(y, a,z, \boldsymbol{\theta})}-1\right] \\[3mm]
			\boldsymbol{j} \left[\dfrac{r}{\pi(y, a, z, \boldsymbol{\theta})}-1\right] 
		   \tag{4.7}
		\end{array}\right),
	\end{equation}
	
	where $\pi(y, a, z, \boldsymbol{\theta}) =f( r =1 \mid y, a, z) = f(r=1 \mid y, a) =\psi\left(\alpha+\beta y+ \gamma a \right)$ in (\ref{equation 4.2}). 
	
	For example, if $z$ is a constant and  there is no other auxiliary variables,  and (\ref{equation 4.7}) is insufficient for estimating unknown $\boldsymbol{\theta} = \left(\alpha, \beta, \gamma \right)^{\mathrm{T}}$. For using the GMM, we need  $q \geq p$. If  there is only a discrete variable $Z$, the requirement $L \geq 2$ is satisfied if $Z$ is not a constant.

	The estimating functions $G(y, r, a, z, \boldsymbol{\theta})$ are motivated by the following equations, if $\boldsymbol{\theta}_0 = (\alpha_0, \beta_0, \gamma_0 )^{\mathrm{T}}$ is the true parameter value,
	\begin{equation} \label{euqation 4.8}
		\begin{aligned}
			E\left[G(y, r, a, z, \boldsymbol{\theta})\right] & = E \left\{\boldsymbol{\eta} \left[\dfrac{r}{ \pi(y, a,z, \boldsymbol{\theta})}-1\right]\right\} \\[3mm]
			& =E\left(E \left\{\boldsymbol{\eta} \left[\dfrac{r}{ \pi(y, a,z, \boldsymbol{\theta})}-1\right] \bigg{|} \  y, a, z\right\}\right) \\[3mm]
			& =E\left\{\boldsymbol{\eta}\left[\dfrac{E(r \mid y, a, z)}{f( r =1 \mid y, a, z)}-1\right]\right\} \\[3mm]
			& =\boldsymbol{0},
		\end{aligned}
		\tag{4.8}
	\end{equation}
	where $\boldsymbol{\eta}= \left(\boldsymbol{k}^{\mathrm{T}}, \boldsymbol{j}^{\mathrm{T}}\right)^{\mathrm{T}}$ is a $q$-dimensional vector. Let $G(y, r, a, z, \boldsymbol{\theta})$ in (\ref{equation 4.7}) be the estimating equations in (\ref{equation 13}), $\widehat{G}_{n}(\boldsymbol{\theta})$ in (\ref{equation 14}) is the sample version. $\widehat{W}$ is given by the two-step method on the above. Then we will obtain $\widehat{\boldsymbol{\theta}} = \left(\hat{\alpha}, \hat{\beta}, \hat{\gamma} \right)^{\mathrm{T}}$ as the two-step GMM estimator of $\boldsymbol{\theta} = \left(\alpha, \beta, \gamma \right)^{\mathrm{T}}$.
	
	\begin{theorem} \label{theorem 2}
		Suppose that   $ E\left[G(y,r,a, z,\boldsymbol{\theta})\right]= \boldsymbol{0}$ only if $\boldsymbol{\theta}=\boldsymbol{\theta}_0$, $\boldsymbol{\theta}_0 \in \Theta$, which is compact and that $E\left[\sup_{\boldsymbol{\theta} \in \Theta}\left\|G(y,r,a, z,\boldsymbol{\theta})\right\|\right]<\infty$. Then, as $n \rightarrow \infty$, $\widehat{\boldsymbol{\theta}} \stackrel{P}{\longrightarrow} \boldsymbol{\theta}_0$  and $\stackrel{P}{\longrightarrow}$ denotes convergence in probability.
		
	\end{theorem}  
	The proof of the Theorem \ref{theorem 2} and more details about GMM can be found in the Supplementary Materials. 
	Let $\nabla_{\boldsymbol{\theta}}(\cdot)$ and $\nabla_{{\boldsymbol{\theta}} {\boldsymbol{\theta}}}(\cdot)$ denote the first- and second-order derivatives with respect to $\boldsymbol{\theta}$.  The asymptotic normality of $\widehat{\boldsymbol{\theta}}$ is established by Theorem \ref{theorem 3} under the following additional Condition \ref{condition 3}.
	\begin{condition} \label{condition 3}
		(i) $\boldsymbol{\theta}_0 \in$ interior of $\Theta$; (ii) $G(y,r,a, z,\boldsymbol{\theta})$ is continuously differentiable in a neighborhood $\mathcal{N}$ of $\boldsymbol{\theta}_0$, with probability approaching one; (iii) $E\left[\|G(y,r,a, z,\boldsymbol{\theta}_0)\|^2\right]$ is finite and  $E\left[\sup _{\boldsymbol{\theta} \in \mathcal{N}}\left\|\nabla_\theta G(y,r,a, z,\boldsymbol{\theta})\right\|\right]<\infty$; (iv)  
		$H = E\left[\nabla_\theta G(y,r,a, z,\boldsymbol{\theta}_0)\right]$ exists and is full of rank.
	\end{condition}
	\begin{theorem} \label{theorem 3}
		Suppose that the assumptions of Theorem \ref{theorem 2} are satisfied, and Condition \ref{condition 3} holds,
		for $\Omega = E[G(y,r,a, z,\boldsymbol{\theta}_0) G(y,r,a, z,\boldsymbol{\theta}_0))^{\mathrm{T}}]$. Then, as $n \rightarrow \infty$, $\sqrt{n}(\widehat{\boldsymbol{\theta}}-\boldsymbol{\theta}_0) \stackrel{\mathscr{L}}{\longrightarrow} N(\boldsymbol{0}, \Delta)$, where $\Delta =\left(H^{\mathrm{T}} W H\right)^{-1} H^{\mathrm{T}} W \Omega W H \left(H^{\mathrm{T}} W H\right)^{-1} $, $\stackrel{\mathscr{L}}{\longrightarrow}$ denotes convergence in distribution.
	\end{theorem}
	The proof of Theorem \ref{theorem 3} can be found in the Supplementary Materials. For asymptotic covariance variance estimation $\widehat{\Delta}$, which can be estimated by substituting estimators for each of $H$, $W$ and $\Omega$. To estimate $\Omega$, we can replace the population moment by a sample average and the true parameter $\boldsymbol{\theta}_0$ by an estimator $\widehat{\boldsymbol{\theta}}$. $\widehat{H}$ is similar to $\widehat{\Omega}$, respectively. To form
	\begin{equation*}
		\widehat{\Omega}= \frac{1}{n} \sum\limits_{i=1}^n G_n(y_i, r_i, a_i, z_i, \widehat{\boldsymbol{\theta}}) G_n(y_i, r_i, a_i, z_i, \widehat{\boldsymbol{\theta}})^{\mathrm{T}} , \quad
		\widehat{H}=\left.\frac{1}{n} \sum_{i=1}^n \frac{\partial G_n(y_i, r_i, a_i, z_i, \boldsymbol{\theta})}{\partial \boldsymbol{\theta}}\right|_{\boldsymbol{\theta}=\widehat{\boldsymbol{\theta}}},
	\end{equation*}
	where $G_n(y_i, r_i, a_i ,   z_i, \boldsymbol{\theta})=\left(g_1\left(y_i, r_i, a_i ,   z_i, \boldsymbol{\theta}\right), \cdots, g_q\left(y_i, r_i, a_i ,   z_i, \boldsymbol{\theta}\right)\right)^{\mathrm{T}}, \quad i=1,\cdots,n, \quad  \boldsymbol{\theta} \in \Theta$. So that we have the Theorem \ref{theorem 4}.
	\begin{theorem} \label{theorem 4}
		If the hypotheses of Theorem \ref{theorem 3} are satisfied, then $\widehat{\Delta} \stackrel{P}{\longrightarrow} \Delta$, where $\widehat=\left(\widehat{H}^{\mathrm{T}} \widehat{W} \widehat{H}\right)^{-1} \widehat{H}^{\mathrm{T}} \widehat{W} \widehat{\Omega} \widehat{W} \widehat{H}\left(\widehat{H}^{\mathrm{T}} \widehat{W} \widehat{H}\right)^{-1}$.
	\end{theorem}
	The proof of Theorem \ref{theorem 4} can be found in the Supplementary Materials. The optimal weight matrix $W_{1} = \Omega^{-1}$, with this choice of $W_{1}$,  $\widehat{W}$ can be constructed by the following, 
	\begin{equation*}
		\widehat{W} = \left\{\frac{1}{n} \sum\limits_{i=1}^n G_n(y_i, r_i, a_i ,   z_i, \tilde{\boldsymbol{\theta}}) G_n(y_i, r_i, a_i ,   z_i, \tilde{\boldsymbol{\theta}})^{\mathrm{T}}\right\}^{-1},
	\end{equation*}
	the asymptotic covariance matrix $\Delta$ reduces to $\left(H^{\mathrm{T}} \Omega H \right)^{-1}$ , and the asymptotic covariance variance estimation $\widehat{\Delta}$ reduces $\left(\widehat{H}^{\mathrm{T}} \widehat{\Omega} \widehat{H} \right)^{-1}$.

	Once $\widehat{\boldsymbol{\theta}}$ is obtained, the mean value of $\mu = E \left[ Y \right]$ can be constructed by the inverse probability weighting with the estimated propensity as the weight function, 
	\begin{equation} \label{equation 4.9}
		\widehat{\mu}=\dfrac{1}{n} \sum_{i=1}^n \dfrac{r_i y_i}{\pi(y_i, a_i ,   z_i, \widehat{\boldsymbol{\theta}})}. \tag{4.9}
	\end{equation} 
	Actually, if  the dimension of the estimating equations $q$ is more than the dimension of the parameters $p$, an additional estimating equation $g_{q+1} = \mu - ry/\pi(y, a, z, \boldsymbol{\theta})$ can be added into the estimating equations $G(y, r, a, z, \boldsymbol{\theta})$ in  (\ref{equation 4.7}). Then $G(y, r, a, z, \boldsymbol{\theta})$ is a set of $q+1 = 2L+1$ estimating equations , and $\boldsymbol{\theta}=\left(\mu, \alpha, \beta, \gamma \right)^{\mathrm{T}}$ is a set of $p+1$ unknown parameters. If $\tilde{\mu}$ is the GMM estimator of the above estimating equations $q+1 = 2L+1$ by the two-step method, as noted by \cite{wang2014instrumental}, the difference between the estimator $\widehat{\mu}$ and $\tilde{\mu}$ is the weighting matrix $\widehat{W}$, the weighting matrix of the esimator $\widehat{\mu}$ is not the optimal and $\tilde{\mu}$ is asymptotically more efficient unless the weighting matrix of the esimator $\widehat{\mu}$ is optimal. But in this paper, we propose $\widehat{\mu}$ to construct the estimator of CACE. The functional delta method can be used to establish the asymptotic results for $\widehat{\mu}$, for simplicity, we omit this part.

	\subsection{Obtaining  CACE from the nonignorable missing outcomes}
	Under  Assumptions \ref{assumption 1}-\ref{assumption 5}, we can estimate CACE for the subpopulation of compliers characterized by $U_i=cp$, by taking the ratio of the average difference in $Y_i$ by instrument and the average difference in $A_i$ by instrument \citep{angrist1996identification}:
	\begin{equation*}\label{equation 4.10}
		E\left[Y_i(1)-Y_i(0) \mid U_i=cp \right]=\frac{E\left[Y_i \mid Z_i=1\right]-E\left[Y_i \mid Z_i=0\right]}{E\left[A_i \mid Z_i=1\right]-E\left[A_i \mid Z_i=0\right]}. \tag{4.10}
	\end{equation*}
     If the outcome $Y$ is fully observed,  CACE can be estimated by equation (\ref{equation 4.10}). So, the missing data problem is equivalent to the problem of estimating the mean difference $E(Y_i \mid Z_i=1)$ and $E(Y_i \mid Z_i=0)$ from incomplete outcome data. 

     In equation (\ref{equation 4.10}), the conditional expectation of the nonignorable missing outcomes $Y$, $E\left[Y_i \mid Z_i=l\right], l=0,1 $ can be estimated by  equation (\ref{equation 4.9}) with the stratification method by the instrument $Z$ ; $E\left[A_i \mid Z_i=l\right], l=0,1$ can be estimated with the fully observed data.  Then the estimator $\widehat{CACE}$ can be obtained by  equation (\ref{equation 4.9}) and (\ref{equation 4.10}).

	\section{Simulations}
	 In this section, we conduct some simulation studies to evaluate the proposed approach in this paper. All the variables are binary variables. We generate the instrument variable, the compliance status variable, the potential outcomes from the following process:
	$$Z  \sim Bernoulli(0.5)$$ 
\begin{table}[htbp]
	\centering
	\caption{True parameters for the compliance status variable $U$}
	\label{Table 1}
	\setlength{\tabcolsep}{5mm}{
	\begin{tabular}{c|ccc}
		\hline
		$U$ & $at$  & $nt$   & $cp$   \\ \hline
		$P$ & $0.2$ & $0.25$ & $0.55$ \\ \hline
	\end{tabular}
}
\end{table}
\begin{table}[htbp]	
	\centering
	\caption{True parameters for the potential outcomes $Y$}
		\label{Table 2}
	\setlength{\tabcolsep}{5mm}{
	\begin{tabular}{c|cc}
		\hline
		$Y(z) \mid U=at$  & $2$   & $4$   \\ 
		$P$               & $0.3$ & $0.7$ \\ \hline
		$Y(z) \mid  U=nt$ & $2$   & $4$   \\ 
		$P$               & $0.7$ & $0.3$ \\ \hline
		$Y(1) \mid U=cp$  & $2$   & $4$   \\ 
		$P$               & $0.4$ & $0.6$ \\ \hline
		$Y(0) \mid U=cp$  & $2$   & $4$   \\ 
		$P$               & $0.6$ & $0.4$ \\ \hline
	\end{tabular}
}
\end{table}

For the outcome $Y$ may be missing not at random, we consider the following logistics missing mechanism in equation (\ref{equation 4.2}):
\begin{equation*} \label{equation 5.1}
	\pi( y, a, z, \boldsymbol{\theta})= P(R=1 \mid Z=z, Y=y, A=a)=\psi(\alpha+\beta y+ \gamma a ) \tag{5.1}
\end{equation*}
where $\psi\left(\alpha+\beta y+ \gamma a  \right) = [1+\exp(-(\alpha+\beta y+\gamma a))]^{-1}$, the true parameters $\boldsymbol{\theta}_0 = (\alpha_0,\beta_0,\gamma_0)^{\mathrm{T}}=(1,-0.1,-0.1)^{\mathrm{T}}$, the missingness indicator $R$ does depend on the outcome $Y$ with missing data itself. The key assumption is Assumption \ref{shadow variable}, which means $R$ is independent of $Z$ conditional on the fully observed variable and possibly unobserved outcome $Y$.  The missing data proportion is approximately between $35\%$ and $40\%$.

From Table \ref{Table 1} and Table \ref{Table 2}, the true value of $CACE$  $=0.4$. We use the approach proposed in Section 4 for our estimation. In Table \ref{Table 3}, the columns with the labels: Bias, SD, and $95\%$ CI represent the average bias for $(\hat{\alpha}-\alpha_0, \hat{\beta}-\beta_0,\hat{\gamma}-\gamma_0,\widehat{CACE}- CACE)$, average standard deviation, and average $95\%$ confidence interval, respectively. We set 1000 replicates under sample sizes 100, 500, 1000, and 2000, respectively. 
\begin{table}[H]	
		\centering
	\caption{ Results of simulation studies}
	\label{Table 3}
	\setlength{\tabcolsep}{5mm}{
	\begin{tabular}{c|cccc}
		\hline
		true value    & n    & Bias    & SD     & $95\%$ CI              \\ \hline
		$\alpha=1.0$  & 100  & 0.0256  & 0.0020 & {[}1.0216, 1.0295{]}   \\
		& 500  & 0.0323  & 0.0048 & {[}1.0229, 1.0417{]}   \\
		& 1000 & 0.0395  & 0.0035 & {[}1.0326, 1.0464{]}   \\
		& 2000 & 0.0373  & 0.0045 & {[}1.0283, 1.0462{]}   \\ \hline
		$\beta=-0.1$  & 100  & 0.0649  & 0.0021 & {[}-0.0392, -0.0308{]} \\
		& 500  & 0.1715  & 0.0075 & {[}0.0567, 0.0863{]}   \\
		& 1000 & 0.1599  & 0.0066 & {[}0.0469, 0.0730{]}   \\
		& 2000 & 0.1561  & 0.0064 & {[}0.0434, 0.0688{]}   \\ \hline
		$\gamma=-0.1$ & 100  & 0.0153  & 0.0021 & {[}-0.0887, -0.0804{]} \\
		& 500  & -0.0188 & 0.0052 & {[}-0.1291, -0.1084{]} \\
		& 1000 & -0.0259 & 0.0052 & {[}-0.1361, -0.1156{]} \\
		& 2000 & -0.0093 & 0.0045 & {[}-0.1183, -0.1004{]} \\ \hline
		$CACE=0.4$    & 100  & -0.0097 & 0.0274 & {[}0.3364, 0.4441{]}   \\
		& 500  & 0.0026  & 0.0110 & {[}0.3809, 0.4243{]}   \\
		& 1000 & -0.0045 & 0.0107 & {[}0.3743, 0.4165{]}   \\
		& 2000 & 0.0013  & 0.0112 & {[}0.3791, 0.4234{]}   \\ \hline
	\end{tabular}
}
\end{table}
 
 From Table \ref{Table 3}, the proposed estimator of  $CACE$ has negligible bias even under the sample size $500$. As the sample size increases, the sd of the $\widehat{CACE}$ becomes much smaller. But the bias of the $\widehat{CACE}$ becomes a nearly stable value when the sample size $n$ equals to $500$ . The estimator $\widehat{\boldsymbol{\theta}}=(\hat{\alpha},\hat{\beta},\hat{\gamma})^{\mathrm{T}}$ all have the small biases and standard deviations even under the sample size $100$. In fact, the proposed estimators of $\widehat{\boldsymbol{\theta}}$ and  $\widehat{CACE}$ in Section 4 all has small biases and standard deviations even under the small sample size $100$. 
 All of the confidence intervals of $\widehat{\boldsymbol{\theta}}$ and  $\widehat{CACE}$ have empirical coverage proportions very close to their nominal values.
	\section{A real data analysis}
    In this section, we apply the proposed estimators to a real data about the political analysis in \cite{esterling2011estimating}. As noted in \cite{esterling2011estimating}, the authors conducted 
    a series of online deliberative field experiments, where current members of the U.S. House of Representatives interacted via a Web-based interface with random samples of their constituents. Some members of Congress conducted the sessions that their constituents participated in. And the constituents interacted with their members in an online chat room. 
    
    An online survey research firm called Knowledge Networks (KN) was responsible for recruiting the constituents from each congressional district and administering the surveys. Each constituent was randomly assigned to three conditions: a deliberative condition that received background reading materials and was asked to complete a survey regarding the background materials(the background materials survey) and to participate in the sessions; an information-only group that only received the background materials and was asked to take the background materials survey; and a true control group. A week after session, KN administered a follow-up survey to subjects in each of the groups. 
    
      There are two questions in the follow-up survey, all subjects were asked ``Please tell us how much you agree or disagree with the following statements'':
     
     1. I don't think public officials care much what people like me think.
     
     2. I have ideas about politics and policy that people in government should listen to.      
    
    The first question is a measure of external efficacy, and the second question is a measure of internal efficacy. In this study, we only consider the first question as the outcome variable: the outcome (\emph{Officials care}) variable  $Y$ equals to $2$ for subjects who ``somewhat disagree'' or ``strongly disagree'' with the first question, and $1$ for subjects who ``somewhat agree'', ``strongly agree'' and ``Neither agree nor disagree''. But in the follow-up survey, some subjects chose not to respond the either question so that the outcome variable $Y$ had missing values. 
    
    \cite{esterling2011estimating} restricted the sample to the $670$ subjects who completed the baseline and background materials surveys and initially indicated a willingness to participate in the deliberative sessions. Thus, the treatment effect compares those who read the background materials and participated in the discussions to those who only read the background materials. Our subject for conducting this experiment is to evaluate  the causal effect of participating in the Online-chat session to Officials care. More details about this deliberative session can be found in \cite{esterling2011estimating}.
    
      The treatment variable $A$ equals to $1$ for subjects who participated in the session and $0$ for subjects who did not participate in the session.  There are 12  pre-treatment (exogenous) variables in \cite{esterling2011estimating}, we choose an exogenous variable as the instrument variable $Z$. The instrument variable $Z$ equals to $1$ for each subject who is able to answer at least four of the ``Delli Carpini and Keeter five''  items correctly on the baseline survey indicating high political knowledge, and $0$ for otherwise. We consider the subject who had high political knowledge might be more likely to participate in the session because they were more likely to show their willingness to participate in the session, which the Assumption \ref{assumption 3} might be reasonable. In fact, we expect the political knowledge is independent of missing mechanism conditional on the participating and possibly missing outcomes, which implies Assumption \ref{shadow variable}. The observed data is reported in Table \ref{Table 4}.
    \begin{table}[H]
    		\centering
    	\caption{ A real data about the political analysis}
    	\label{Table 4}
    	\setlength{\tabcolsep}{1mm}{
\begin{tabular}{c|c|c|c|c}
	\hline
	& $Z=1$, $A=1$ & $Z=1$, $A=0$ & $Z=0$, $A=1$ & $Z=0$, $A=0$ \\ \hline
	$R=1$, $Y=1$ & 130          & 139         & 62          & 82          \\ \hline
	$R=1$, $Y=2$ & 67           & 24          & 12          & 11          \\ \hline
	$R=0$, $Y=.$ & 21           & 72          & 5           & 45          \\ \hline
\end{tabular}
    }
    \end{table}
    The samples size $n$ equals to $670$ with $143$ missing values. The missing proportion is about $21.34\%$(about $143/670 \approx 0.2134$). In the treatment group ($A=1$) who participated in the session, the missing proportion is about $8.75\%$($26/297\approx 0.0875$), but in the control group ($A=0$) who did not participate in the session, the missing proportion is about $31.36\%$($117/373 \approx 0.3136$). Those who participated in the session were more likely to respond to the follow-up survey than those not. We consider the missing mechanism was influenced by  the subject whether  participated in the session. The missing mechanism $R$ may possibly depend on the missing outcomes $Y$ because of the questions in the follow-up survey. Hence it is possible that the outcomes are missing not at random. 
    
    The estimates of $\boldsymbol{\theta}$ and $CACE$ are shown in Table \ref{Table 5}. The columns of Table \ref{Table 5} correspond to the parameters, point estimate, Standard Deviation and  $95\%$  Confidence Interval, respectively. The assumption about the missing mechanism is similar to the equation (\ref{equation 5.1}) in Section 5.
   
     \begin{table}[H]
 	\centering
 	\caption{Results for the estimates of Parameters and $CACE$}
 	\label{Table 5}
 	\setlength{\tabcolsep}{5mm}{
    	\begin{tabular}{c|c|c|c}
    		\hline
    		Patameters     & Estimate &       SD   & $95\%$  Confidence Interval \\ \hline
    		$\hat{\alpha}$   & 1.6204     & 0.0349 & {[}1.5519, 1.6888{]}        \\ \hline
    		$\hat{\beta}$    & -0.2225    & 0.0234 & {[}-0.2684, -0.1766{]}      \\ \hline
    		$\hat{\gamma}$   & 0.1249     & 0.0249 & {[}0.0759, 0.1739{]}        \\ \hline
    		$\widehat{CACE}$ & 1.3234     & 0.0190 & {[}1.2861, 1.3608{]}        \\ \hline
    	\end{tabular}
    }
    \end{table}
    From Table \ref{Table 5}, we can see that, under the assumption of the missing mechanism in equation (\ref{equation 5.1}), the sign of  $\hat{\beta}$ is negative and the sign of $\hat{\gamma}$ is positive, which means the more disagreement on the first questions in the follow-up survey, the more likely the outcome variable $Y$ is to be missing and for those who did not participate in the session, their outcomes might be more likely to have missing values. The estimate of CACE is 1.3234, which means participating in the session would propose a  positive  effect on Officials care significantly, which means a subject who participated in the session might change their opinions from agreement to disagreement about the first question. Our results are similar to \cite{esterling2011estimating}, who also produced a significant result.

    \section{Discussions}
     How to obtain the average causal effects from the nonignorable missing data is a challenging problem. In fact, under the principal stratification framework, it can not be estimated the average causal effects for  all population, we can only estimate the complier average causal effects. But in a real situation, we can not observe the compliance status of an unit. With the nonignorable missing outcomes $Y$, the identification of  CACE can not be established without additional assumptions. Some different missing data mechanisms are discussed in the previous work.
     
     In this paper, we compare LI and ODN missing data mechanisms with the shadow variable assumption. And then we establish the nonparametric identification about the full data distribution from the nonignorable missing outcomes with a shadow variable. However, it might not be easy to find such a variable in the real analysis. Some Prior knowledge is usually needed in practice.
     For inference, we impose a parametric assumption on the missing data mechanism  and no assumption on the outcome model. With the GMM method, we estimate the parameters in the propensity. And then we establish some results of the consistency and the asymptotic properties of the parameters. The next, we recover the mean value of the outcome $Y$ and  estimate  CACE by equation (\ref{equation 4.10}). In the discussions, we discuss a few possible future generalizations of the proposed approach in this paper.
   \subsection{Adjustment for measured confounding with a covariate $V$}
     When a discrete covariate $V$ can be fully observed, we adjust the measured confounding with a stratification method. Some assumptions will be established conditional on $V$.
     
    Randomization Assumption \ref{assumption 2} is replaced by : For given  $V$,  $(Y(0), Y(1), A(0), A(1))$ is independent of $Z$ ($Z \perp \!\!\! \perp (Y(0), Y(1), A(0), A(1)) \mid V$).
     Shadow variable Assumption \ref{shadow variable} is replaced by:	(a'): $Z \perp \!\!\! \perp R \mid(Y, V, A)$;   (b'): $Z \not \perp \!\!\! \perp Y \mid(R=1, V, A)$.  Under the assumptions conditional on $V$ and other essential assumptions, the Definition \ref{complier average causal effect} of CACE is replaced by CACE$=E\left[Y_i(1)-Y_i(0) \mid U_i=cp, V_i=v_i\right]$.

      We present a figure to help readers to understand the framework of this paper with a covariate $V$.
     
     \begin{center}
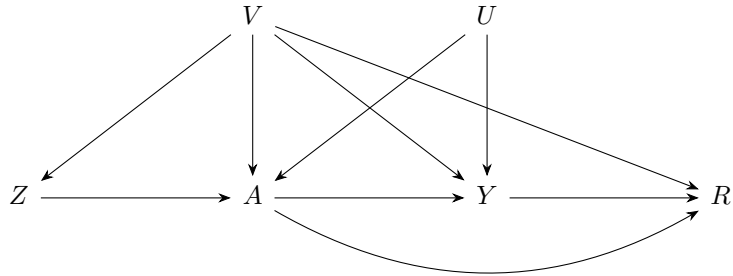

     	\begin{tikzcd}[ampersand replacement=\&,row sep=large,column sep=large]
     		\&  \& V  \arrow[lldd,->,>=Stealth] \arrow[dd,->,>=Stealth] \arrow[rrdd,->,>=Stealth] \arrow[rrrrdd,->,>=Stealth] \&  \& U \arrow[lldd,->,>=Stealth] \arrow[dd,->,>=Stealth] \&  \&   \\
     		\&  \&                                                                  \&  \&                           \&  \&   \\
     		Z \arrow[rr,->,>=Stealth] \&  \& A \arrow[rrrr,->,>=Stealth,bend right=30] \arrow[rr,->,>=Stealth]                                                     \&  \& Y \arrow[rr,->,>=Stealth]              \&  \& R
     	\end{tikzcd}
     \end{center}
     \captionof{figure}{A directed acyclic graph model for this paper with a covariate $V$}	
     And then CACE can be estimated by the following equation:
     	\begin{equation*}
     	E\left[Y_i(1)-Y_i(0) \mid U_i=cp, V_i=v_i\right]=\frac{E\left[Y_i \mid Z_i=1,V_i=v_i\right]-E\left[Y_i \mid Z_i=0,V_i=v_i\right]}{E\left[A_i \mid Z_i=1,V_i=v_i\right]-E\left[A_i \mid Z_i=0,V_i=v_i\right]}. 
     \end{equation*}
     \subsection{How to obtain the average causal effects within all population}
       \cite{wang2018bounded} discussed two framework of the instrument variable. Under the linear  structural  equation models (SEMs), which implies the effect homogeneity, the average causal effects can be identified and estimated. But the  structural  equation models are so sensitive to the model assumptions. In this paper, we impose no assumption on the outcome model, including the identification and estimation.  Probably imposing some assumptions on the outcome model might be able to estimate the average causal effects from the nonignorable missing outcomes under some new missing data mechanisms. 

	\newpage 
	\bibliographystyle{apalike}	
	\bibliography{document.bib}

\begin{thebibliography}{}

\bibitem[Angrist and Imbens, 1995]{angrist1995two}
Angrist, J.~D. and Imbens, G.~W. (1995).
\newblock Two-stage least squares estimation of average causal effects in
  models with variable treatment intensity.
\newblock {\em Journal of the American statistical Association},
  90(430):431--442.

\bibitem[Angrist et~al., 1996]{angrist1996identification}
Angrist, J.~D., Imbens, G.~W., and Rubin, D.~B. (1996).
\newblock Identification of causal effects using instrumental variables.
\newblock {\em Journal of the American statistical Association},
  91(434):444--455.

\bibitem[Baker and Laird, 1988]{baker1988regression}
Baker, S.~G. and Laird, N.~M. (1988).
\newblock Regression analysis for categorical variables with outcome subject to
  nonignorable nonresponse.
\newblock {\em Journal of the American Statistical association},
  83(401):62--69.

\bibitem[Bang and Robins, 2005]{bang2005doubly}
Bang, H. and Robins, J.~M. (2005).
\newblock Doubly robust estimation in missing data and causal inference models.
\newblock {\em Biometrics}, 61(4):962--973.

\bibitem[Chen et~al., 2015]{chen2015semiparametric}
Chen, H., Ding, P., Geng, Z., and Zhou, X.-H. (2015).
\newblock Semiparametric inference of the complier average causal effect with
  nonignorable missing outcomes.
\newblock {\em ACM Transactions on Intelligent Systems and Technology (TIST)},
  7(2):1--15.

\bibitem[Chen et~al., 2009]{chen2009identifiability}
Chen, H., Geng, Z., and Zhou, X.-H. (2009).
\newblock Identifiability and estimation of causal effects in randomized trials
  with noncompliance and completely nonignorable missing data.
\newblock {\em Biometrics}, 65(3):675--682.

\bibitem[Chen, 2003]{chen2003note}
Chen, H.~Y. (2003).
\newblock A note on the prospective analysis of outcome-dependent samples.
\newblock {\em Journal of the Royal Statistical Society: Series B (Statistical
  Methodology)}, 65(2):575--584.

\bibitem[Chen, 2004]{chen2004nonparametric}
Chen, H.~Y. (2004).
\newblock Nonparametric and semiparametric models for missing covariates in
  parametric regression.
\newblock {\em Journal of the American Statistical Association},
  99(468):1176--1189.

\bibitem[Chen, 2007]{yun2007semiparametric}
Chen, H.~Y. (2007).
\newblock A semiparametric odds ratio model for measuring association.
\newblock {\em Biometrics}, 63(2):413--421.

\bibitem[Darolles et~al., 2011]{darolles2011nonparametric}
Darolles, S., Fan, Y., Florens, J.-P., and Renault, E. (2011).
\newblock Nonparametric instrumental regression.
\newblock {\em Econometrica}, 79(5):1541--1565.

\bibitem[Ding and Geng, 2014]{ding2014identifiability}
Ding, P. and Geng, Z. (2014).
\newblock Identifiability of subgroup causal effects in randomized experiments
  with nonignorable missing covariates.
\newblock {\em Statistics in Medicine}, 33(7):1121--1133.

\bibitem[Ding and Li, 2018]{ding2018causal}
Ding, P. and Li, F. (2018).
\newblock Causal inference: A missing data perspective.
\newblock {\em Statistical Science}, 33(2):214--237.

\bibitem[d’Haultfoeuille, 2010]{d2010new}
d’Haultfoeuille, X. (2010).
\newblock A new instrumental method for dealing with endogenous selection.
\newblock {\em Journal of Econometrics}, 154(1):1--15.

\bibitem[Esterling et~al., 2011]{esterling2011estimating}
Esterling, K.~M., Neblo, M.~A., and Lazer, D.~M. (2011).
\newblock Estimating treatment effects in the presence of noncompliance and
  nonresponse: The generalized endogenous treatment model.
\newblock {\em Political Analysis}, 19(2):205--226.

\bibitem[Frangakis and Rubin, 1999]{frangakis1999addressing}
Frangakis, C.~E. and Rubin, D.~B. (1999).
\newblock Addressing complications of intention-to-treat analysis in the
  combined presence of all-or-none treatment-noncompliance and subsequent
  missing outcomes.
\newblock {\em Biometrika}, 86(2):365--379.

\bibitem[Frangakis and Rubin, 2002]{frangakis2002principal}
Frangakis, C.~E. and Rubin, D.~B. (2002).
\newblock Principal stratification in causal inference.
\newblock {\em Biometrics}, 58(1):21--29.

\bibitem[Goldberger, 1972]{goldberger1972structural}
Goldberger, A.~S. (1972).
\newblock Structural equation methods in the social sciences.
\newblock {\em Econometrica: Journal of the Econometric Society}, 40:979--1001.

\bibitem[Greenlees et~al., 1982]{greenlees1982imputation}
Greenlees, J.~S., Reece, W.~S., and Zieschang, K.~D. (1982).
\newblock Imputation of missing values when the probability of response depends
  on the variable being imputed.
\newblock {\em Journal of the American Statistical Association},
  77(378):251--261.

\bibitem[Hall, 2005]{hall2004generalized}
Hall, A.~R. (2005).
\newblock {\em Generalized method of moments}.
\newblock Oxford: Oxford University Press.

\bibitem[Hansen, 1982]{hansen1982large}
Hansen, L.~P. (1982).
\newblock Large sample properties of generalized method of moments estimators.
\newblock {\em Econometrica: Journal of the econometric society}, pages
  1029--1054.

\bibitem[Higgins et~al., 2008]{higgins2008imputation}
Higgins, J.~P., White, I.~R., and Wood, A.~M. (2008).
\newblock Imputation methods for missing outcome data in meta-analysis of
  clinical trials.
\newblock {\em Clinical trials}, 5(3):225--239.

\bibitem[Horvitz and Thompson, 1952]{horvitz1952generalization}
Horvitz, D.~G. and Thompson, D.~J. (1952).
\newblock A generalization of sampling without replacement from a finite
  universe.
\newblock {\em Journal of the American statistical Association},
  47(260):663--685.

\bibitem[Imai, 2009]{imai2009statistical}
Imai, K. (2009).
\newblock Statistical analysis of randomized experiments with non-ignorable
  missing binary outcomes: an application to a voting experiment.
\newblock {\em Journal of the Royal Statistical Society: Series C (Applied
  Statistics)}, 58(1):83--104.

\bibitem[Imbens and Angrist, 1994]{imbens1994identification}
Imbens, G.~W. and Angrist, J.~D. (1994).
\newblock Identification and estimation of local average treatment effects.
\newblock {\em Econometrica}, 62(2):467--475.

\bibitem[Kim and Yu, 2011]{kim2011semiparametric}
Kim, J.~K. and Yu, C.~L. (2011).
\newblock A semiparametric estimation of mean functionals with nonignorable
  missing data.
\newblock {\em Journal of the American Statistical Association},
  106(493):157--165.

\bibitem[Kott, 2014]{kott2014calibration}
Kott, P.~S. (2014).
\newblock Calibration weighting when model and calibration variables can
  differ.
\newblock In {\em Contributions to sampling statistics}, pages 1--18. Springer.

\bibitem[Li and Zhou, 2017]{li2017identifiability}
Li, W. and Zhou, X.-H. (2017).
\newblock Identifiability and estimation of causal mediation effects with
  missing data.
\newblock {\em Statistics in Medicine}, 36(25):3948--3965.

\bibitem[Little, 1993]{little1993pattern}
Little, R.~J. (1993).
\newblock Pattern-mixture models for multivariate incomplete data.
\newblock {\em Journal of the American Statistical Association},
  88(421):125--134.

\bibitem[Little, 1994]{little1994class}
Little, R.~J. (1994).
\newblock A class of pattern-mixture models for normal incomplete data.
\newblock {\em Biometrika}, 81(3):471--483.

\bibitem[Little and Rubin, 2002]{littlerubin2002}
Little, R.~J. and Rubin, D.~B. (2002).
\newblock {\em Statistical analysis with missing data}.
\newblock Wiley: New York.

\bibitem[Miao et~al., 2016]{miao2016identifiability}
Miao, W., Ding, P., and Geng, Z. (2016).
\newblock Identifiability of normal and normal mixture models with nonignorable
  missing data.
\newblock {\em Journal of the American Statistical Association},
  111(516):1673--1683.

\bibitem[Miao et~al., 2019]{miao2015identification}
Miao, W., Liu, L., Tchetgen, E.~T., and Geng, Z. (2019).
\newblock Identification, doubly robust estimation, and semiparametric
  efficiency theory of nonignorable missing data with a shadow variable.
\newblock {\em arXiv preprint arXiv:1509.02556v3}.

\bibitem[Miao and Tchetgen, 2018]{miao2018identification}
Miao, W. and Tchetgen, E.~T. (2018).
\newblock Identification and inference with nonignorable missing covariate
  data.
\newblock {\em Statistica Sinica}, 28(4):2049--2067.

\bibitem[Miao and Tchetgen~Tchetgen, 2016]{miao2016varieties}
Miao, W. and Tchetgen~Tchetgen, E.~J. (2016).
\newblock On varieties of doubly robust estimators under missingness not at
  random with a shadow variable.
\newblock {\em Biometrika}, 103(2):475--482.

\bibitem[Newey and McFadden, 1994]{newey1994large}
Newey, W.~K. and McFadden, D. (1994).
\newblock Large sample estimation and hypothesis testing.
\newblock {\em Handbook of econometrics}, 4:2111--2245.

\bibitem[Newey and Powell, 2003]{newey2003instrumental}
Newey, W.~K. and Powell, J.~L. (2003).
\newblock Instrumental variable estimation of nonparametric models.
\newblock {\em Econometrica}, 71(5):1565--1578.

\bibitem[Neyman, 1923]{neyman1923application}
Neyman, J.~S. (1923).
\newblock On the application of probability theory to agricultural experiments.
  essay on principles. section 9.(translated and edited by dm dabrowska and tp
  speed, statistical science (1990), 5, 465-480).
\newblock {\em Annals of Agricultural Sciences}, 10:1--51.

\bibitem[Niu et~al., 2014]{niu2014empirical}
Niu, C., Guo, X., Xu, W., and Zhu, L. (2014).
\newblock Empirical likelihood inference in linear regression with nonignorable
  missing response.
\newblock {\em Computational Statistics \& Data Analysis}, 79:91--112.

\bibitem[O'Malley and Normand, 2005]{o2005likelihood}
O'Malley, A.~J. and Normand, S.-L.~T. (2005).
\newblock Likelihood methods for treatment noncompliance and subsequent
  nonresponse in randomized trials.
\newblock {\em Biometrics}, 61(2):325--334.

\bibitem[Osius, 2004]{osius2004association}
Osius, G. (2004).
\newblock The association between two random elements: A complete
  characterization and odds ratio models.
\newblock {\em Metrika}, 60(3):261--277.

\bibitem[Qin et~al., 2002]{qin2002estimation}
Qin, J., Leung, D., and Shao, J. (2002).
\newblock Estimation with survey data under nonignorable nonresponse or
  informative sampling.
\newblock {\em Journal of the American Statistical Association},
  97(457):193--200.

\bibitem[Robins and Ritov, 1997]{robins1997toward}
Robins, J.~M. and Ritov, Y. (1997).
\newblock Toward a curse of dimensionality appropriate (coda) asymptotic theory
  for semi-parametric models.
\newblock {\em Statistics in medicine}, 16(3):285--319.

\bibitem[Rosenbaum and Rubin, 1983]{rosenbaum1983central}
Rosenbaum, P.~R. and Rubin, D.~B. (1983).
\newblock The central role of the propensity score in observational studies for
  causal effects.
\newblock {\em Biometrika}, 70(1):41--55.

\bibitem[Roy, 2003]{roy2003modeling}
Roy, J. (2003).
\newblock Modeling longitudinal data with nonignorable dropouts using a latent
  dropout class model.
\newblock {\em Biometrics}, 59(4):829--836.

\bibitem[Rubin, 1973]{rubin1973matching}
Rubin, D.~B. (1973).
\newblock Matching to remove bias in observational studies.
\newblock {\em Biometrics}, 29:159--183.

\bibitem[Rubin, 1974]{rubin1974estimating}
Rubin, D.~B. (1974).
\newblock Estimating causal effects of treatments in randomized and
  nonrandomized studies.
\newblock {\em Journal of educational Psychology}, 66(5):688--701.

\bibitem[Rubin, 1976]{rubin1976inference}
Rubin, D.~B. (1976).
\newblock Inference and missing data.
\newblock {\em Biometrika}, 63(3):581--592.

\bibitem[Rubin, 1979]{rubin1979using}
Rubin, D.~B. (1979).
\newblock Using multivariate matched sampling and regression adjustment to
  control bias in observational studies.
\newblock {\em Journal of the American Statistical Association},
  74(366a):318--328.

\bibitem[Rubin, 1980]{rubin1980randomization}
Rubin, D.~B. (1980).
\newblock Randomization analysis of experimental data: The fisher randomization
  test comment.
\newblock {\em Journal of the American statistical association},
  75(371):591--593.

\bibitem[Shao and Wang, 2016]{shao2016semiparametric}
Shao, J. and Wang, L. (2016).
\newblock Semiparametric inverse propensity weighting for nonignorable missing
  data.
\newblock {\em Biometrika}, 103(1):175--187.

\bibitem[Stuart, 2010]{stuart2010matching}
Stuart, E.~A. (2010).
\newblock Matching methods for causal inference: A review and a look forward.
\newblock {\em Statistical science: a review journal of the Institute of
  Mathematical Statistics}, 25(1):1--21.

\bibitem[Tang et~al., 2003]{tang2003analysis}
Tang, G., Little, R.~J., and Raghunathan, T.~E. (2003).
\newblock Analysis of multivariate missing data with nonignorable nonresponse.
\newblock {\em Biometrika}, 90(4):747--764.

\bibitem[Tang et~al., 2014]{tang2014empirical}
Tang, N., Zhao, P., and Zhu, H. (2014).
\newblock Empirical likelihood for estimating equations with nonignorably
  missing data.
\newblock {\em Statistica Sinica}, 24:723--747.

\bibitem[Tauchen, 1985]{tauchen1985diagnostic}
Tauchen, G. (1985).
\newblock Diagnostic testing and evaluation of maximum likelihood models.
\newblock {\em Journal of Econometrics}, 30(1-2):415--443.

\bibitem[Wang and Tchetgen~Tchetgen, 2018]{wang2018bounded}
Wang, L. and Tchetgen~Tchetgen, E. (2018).
\newblock Bounded, efficient and multiply robust estimation of average
  treatment effects using instrumental variables.
\newblock {\em Journal of the Royal Statistical Society Series B: Statistical
  Methodology}, 80(3):531--550.

\bibitem[Wang et~al., 2014]{wang2014instrumental}
Wang, S., Shao, J., and Kim, J.~K. (2014).
\newblock An instrumental variable approach for identification and estimation
  with nonignorable nonresponse.
\newblock {\em Statistica Sinica}, 24:1097--1116.

\bibitem[Wright, 1928]{wright1928tariff}
Wright, P.~G. (1928).
\newblock {\em Tariff on animal and vegetable oils}.
\newblock Macmillan Company, New York.

\bibitem[Wu and Carroll, 1988]{wu1988estimation}
Wu, M.~C. and Carroll, R.~J. (1988).
\newblock Estimation and comparison of changes in the presence of informative
  right censoring by modeling the censoring process.
\newblock {\em Biometrics}, 44:175--188.

\bibitem[Yang et~al., 2019]{yang2019causal}
Yang, S., Wang, L., and Ding, P. (2019).
\newblock Causal inference with confounders missing not at random.
\newblock {\em Biometrika}, 106(4):875--888.

\bibitem[Zhang et~al., 2018]{zhang2018generalized}
Zhang, L., Lin, C., and Zhou, Y. (2018).
\newblock Generalized method of moments for nonignorable missing data.
\newblock {\em Statistica Sinica}, 28(4):2107--2124.

\bibitem[Zhao et~al., 2013]{zhao2013empirical}
Zhao, H., Zhao, P.~Y., and Tang, N.~S. (2013).
\newblock Empirical likelihood inference for mean functionals with nonignorably
  missing response data.
\newblock {\em Computational Statistics \& Data Analysis}, 66:101--116.

\bibitem[Zhao and Shao, 2015]{zhao2015semiparametric}
Zhao, J. and Shao, J. (2015).
\newblock Semiparametric pseudo-likelihoods in generalized linear models with
  nonignorable missing data.
\newblock {\em Journal of the American Statistical Association},
  110(512):1577--1590.

\bibitem[Zhou and Li, 2006]{zhou2006itt}
Zhou, X.-H. and Li, S.~M. (2006).
\newblock Itt analysis of randomized encouragement design studies with missing
  data.
\newblock {\em Statistics in medicine}, 25(16):2737--2761.

\end{thebibliography}
	
	\addcontentsline{toc}{section}{References}	
	
	\newpage

	\begin{center}
		\section*{\bf \huge Supplementary Materials} 
	\end{center}
	\addcontentsline{toc}{section}{Supplementary Materials}
	
	\bigskip 
	\subsection*{\bf Appendix A: Proofs of the propositions}
	\addcontentsline{toc}{subsection}{Appendix A: Proofs of the propositions} 	
	Throughout the paper, $\mathcal{P}(\cdot)$ denotes  the counting measure for a discrete variable.
	\subsubsection*{Proofs of the proposition \ref{proposition 1}}
	\addcontentsline{toc}{subsubsection}{Proofs of the proposition \ref{proposition 1}}
	Under Assumption \ref{shadow variable}, we have that for all $(A,Y,Z)$
	\begin{equation*}
		\operatorname{OR}(a, y, z)=\operatorname{OR}(a, y) \equiv \frac{f(r=0 \mid a, y) f(r=1 \mid a, y=1)}{f(r=1 \mid a, y) f(r=0 \mid a, y=1) }, \tag{4}
	\end{equation*}
	\vskip.1in
	\begin{equation*}
		\begin{gathered}
			f(y, r \mid a, z)=c(a, z) f(r \mid a, y=1) f(y \mid r=1, a, z)\{\operatorname{OR}(a, y)\}^{1-r},\\[2mm] \tag{5}
			c(a, z)=\frac{f(r=1 \mid a)}{f(r=1 \mid a, y=1)} \frac{f(z \mid r=1, a)}{f(z \mid a)},
		\end{gathered}
	\end{equation*}
	\vskip.1in
	\begin{equation*}
		f(r=1 \mid a, y=1)=\frac{E\left[\operatorname{OR}(a, y) \mid r=1, a\right]}{f(r=0 \mid a) / f(r=1 \mid a)+E\left[\operatorname{OR}(a, y) \mid r=1, a\right]}.  \tag{6}
	\end{equation*}
	
	\paragraph*{Proof of equation (\ref{equation 4})}
	\addcontentsline{toc}{paragraph}{Proof of equation (\ref{equation 4})}   
	Given Assumption \ref{shadow variable} and  according to equation (\ref{equation 3}), we have the following results
	\begin{equation*}
		\begin{aligned}
			\operatorname{OR}(a, y, z)&=\frac{f(y \mid r=0, a, z) f(y=1 \mid r=1, a, z)}{f(y \mid r=1, a, z) f(y=1 \mid r=0, a, z)}\\[3mm]
			& =\dfrac{\dfrac{f(y, r=0, a, z)}{f(r=0, a, z)}  \dfrac{f(y=1, r=1, a, z)}{f(r=1, a, z)}}{\dfrac{f(y, r=1, a, z)}{f(r=1, a, z)}  \dfrac{f(y=1, r=0, a, z)}{f(r=0, a, z)}} \\[3mm]
			& =\frac{f(y, r=0, a, z)  f(y=1, r=1, a, z)}{f(y, r=1, a, z)  f(y=1, r=0, a, z)}\\[3mm]
			&=\dfrac{\dfrac{f(y, r=0, a, z)}{f(a, y)} \dfrac{f(y=1, r=1, a, z)}{f(a, y=1)}}{\dfrac{f(y, r=1, a, z)}{f(a, y)} \dfrac{f(y=1, r=0, a, z)}{f(a, y=1)}}\\[3mm]
			&=\frac{f(r=0, z \mid a, y)  f(r=1, z \mid a, y=1)}{f(r=1, z \mid a, y)  f(r=0, z \mid a, y=1)}\\[3mm]
			&= \frac{f(r=0 \mid a, y) f(r=1 \mid a, y=1)}{f(r=1 \mid a, y) f(r=0 \mid a, y=1) } \equiv \operatorname{OR}(a, y) 
		\end{aligned}
	\end{equation*}
	
	\paragraph*{Proof of equation (\ref{equation 5})}
	\addcontentsline{toc}{paragraph}{Proof of equation (\ref{equation 5})} 
	To proof equation (\ref{equation 5}), we prove that for $R=1$ and $R=0$, respectively.
	\vskip.05in
	\noindent {\it When $R=1$, we have that}
	\begin{equation*}
		\begin{aligned}
			\text{right side of the equation (\ref{equation 5})} &= c(a, z) f(r=1 \mid a, y=1) f(y \mid r=1, a, z) \\[3mm]
			&=\dfrac{f(r=1 \mid a)}{f(r=1 \mid a, y=1)} \dfrac{f(z \mid r=1, a)}{f(z \mid a)} f(r=1 \mid a, y=1) f(y \mid r=1, a, z)\\[3mm]
			&=f(r=1 \mid a) \dfrac{f(z \mid r=1, a)}{f(z \mid a)} f(y \mid r=1, a, z) \\[3mm]
			&=\dfrac{f(r=1, a)}{f(a)} \dfrac{\dfrac{f(z, r=1, a)}{f(r=1, a)}}{\dfrac{f(z, a)}{f(a)}} \dfrac{f(y, r=1, a, z)}{f(r=1, a, z)} \\[3mm]
			& =\dfrac{f(y, r=1, a, z)}{f(z, a)}=f(y, r=1 \mid a, z)=  \text{left side of the equation (\ref{equation 5})}
		\end{aligned}
	\end{equation*}
	\vskip.05in
	\noindent {\it When $R=0$, from equation (\ref{equation 4}) and Assumption \ref{shadow variable}, we have that}
	\begin{equation*}
		\begin{aligned}
			\text{right side of the equation (\ref{equation 5})} = &c(a, z) f(r=0 \mid a, y=1) f(y \mid r=1, a, z) {\operatorname{OR}}(a, y)  \\[3mm]
			=&\dfrac{f(r=1 \mid a)}{f(r=1 \mid a, y=1)} \dfrac{f(z \mid r=1, a)}{f(z \mid a)} f(r=0 \mid a, y=1) f(y \mid r=1, a, z) \\[3mm]
			& \dfrac{f(r=0 \mid a, y)}{f(r=1 \mid a, y)} \dfrac{f(r=1 \mid a, y=1)}{f(r=0 \mid a, y=1)}\\[3mm]
			=	& f(r=1 \mid a) \dfrac{f(z \mid r=1, a)}{f(z \mid a)} f(y \mid r=1, a, z) \dfrac{f(r=0 \mid a, y)}{f(r=1 \mid a, y)} \\[3mm]
			=	&\dfrac{f(r=1, a)}{f(a)} \dfrac{\dfrac{f(z, r=1, a)}{f(r=1, a)}}{\dfrac{f(z, a)}{f(a)}} \dfrac{f(y, r=1, a, z)}{f(r=1, a, z)} \dfrac{f(r=0, z \mid a, y)}{f(r=1, z \mid a, y)} \\[3mm]
			=	&\dfrac{f(z, r=1, a)}{f(z, a)}  \dfrac{f(y, r=1, a, z)}{f(r=1, a, z)}  \frac{f(r=0, z, a, y)}{f(r=1, z, a, y)}\\[3mm]
			=& \dfrac{f(r=0, y, z, a)}{f(z, a)} = f(y, r=0 \mid a, z) = \text{left side of the equation (\ref{equation 5})}
		\end{aligned}
	\end{equation*}
	In conclusion, equation (\ref{equation 5}) holds for the value $R=1$ and $R=0$.
	
	\paragraph*{Proof of equation (\ref{equation 6})}
	\addcontentsline{toc}{paragraph}{Proof of equation (\ref{equation 6})}
	From equation (\ref{equation 4}), we notice that the odds ratio function $\operatorname{OR}(a, y)$ is the function of the variables $(A,Y)$, by the property of conditional mathematical expectation, we have that
	
	\begin{equation*}
		\begin{aligned}
			& E\left[\operatorname{OR}(a, y) \mid r=1, a\right] \\[3mm]
			& =\int \dfrac{f(r=0 \mid a, y) f(r=1 \mid a, y=1)}{f(r=1 \mid a, y) f(r=0 \mid a, y=1)} f(y \mid r=1, a) d \mathcal{P}(y)\\[3mm]
			&=\dfrac{f(r=1 \mid a, y=1)}{f(r=0 \mid a, y=1)} \int \dfrac{f(r=0 \mid a, y)}{f(r=1 \mid a, y)} f(y \mid r=1, a) d \mathcal{P}(y)\\[3mm]
			&=\dfrac{f(r=1 \mid a, y=1)}{f(r=0 \mid a, y=1)} \int \dfrac{f(r=0, a, y)}{f(r=1, a, y)} \dfrac{f(y, r=1, a)}{f(r=1, a)} d \mathcal{P}(y)\\[3mm]
			&=\dfrac{f(r=1 \mid a, y=1)}{f(r=0 \mid a, y=1)} \int \dfrac{f(r=0, a, y)}{f(r=1, a)} d \mathcal{P}(y)\\[3mm]
			& =\dfrac{f(r=1 \mid a, y=1)}{f(r=0 \mid a, y=1)}  \dfrac{1}{f(r=1, a)} \int f(r=0, a, y) d\mathcal{P}(y)\\[3mm]
			& =\dfrac{f(r=1 \mid a, y=1)}{f(r=0 \mid a, y=1)} \frac{f(r=0, a)}{f(r=1, a)}
		\end{aligned}
	\end{equation*}
	
	\begin{equation*}
		\begin{aligned}
			\text{right side of the equation (\ref{equation 6})} &=\dfrac{E\left[\operatorname{OR}(a, y) \mid r=1, a\right]}{f(r=0 \mid a) / f(r=1 \mid a)+E\left[\operatorname{OR}(a, y) \mid r=1, a\right]}\\[3mm]
			&=\dfrac{\dfrac{f(r=1 \mid a, y=1)}{f(r=0 \mid a, y=1)} \dfrac{f(r=0, a)}{f(r=1, a)}}{\dfrac{f(r=0 \mid a)}{f(r=1 \mid a)}+\dfrac{f(r=1 \mid a, y=1)}{f(r=0 \mid a, y=1)} \dfrac{f(r=0, a)}{f(r=1, a)}}\\[3mm]
			& =\dfrac{\dfrac{f(r=1 \mid a, y=1)}{f(r=0 \mid a, y=1)} \dfrac{f(r=0, a)}{f(r=1, a)} }{\dfrac{f(r=0, a)}{f(r=1, a)}+\dfrac{f(r=1 \mid a, y=1)}{f(r=0 \mid a, y=1)} \dfrac{f(r=0, a)}{f(r=1, a)} }\\[3mm]
			& =\dfrac{\dfrac{f(r=1 \mid a, y=1)}{f(r=0 \mid a, y=1)}}{1+\dfrac{f(r=1 \mid a, y=1)}{f(r=0 \mid a, y=1)}} \\[3mm]
			& =\dfrac{\dfrac{f(r=1 \mid a, y=1)}{f(r=0 \mid a, y=1)}}{\dfrac{f(r=0 \mid a, y=1)+f(r=1 \mid a, y=1)}{f(r=0, a, y=1)}} \\[3mm]
			& = f(r=1 \mid a, y=1) = \text{left side of the equation (\ref{equation 6})}
		\end{aligned}
	\end{equation*}
	
	\newpage
	\subsubsection*{Proofs of the proposition \ref{proposition 2}}
	\addcontentsline{toc}{subsubsection}{Proofs of the proposition \ref{proposition 2}}
	Under Assumption \ref{shadow variable}, we have that
	\begin{equation*}
		\begin{aligned}
			f(r=1 \mid a, y) & =f(r=1 \mid a, y, z),  \\[2mm] 
			& =\dfrac{f(r=1 \mid a, y=1)}{f(r=1 \mid a, y=1)+\operatorname{OR}(a, y) f(r=0 \mid a, y=1)}, 
		\end{aligned}
		\tag{7}
	\end{equation*}
	\vskip.1in
	\begin{equation*} 
		f(y \mid r=0, a, z)=\dfrac{\operatorname{OR}(a, y) f(y \mid r=1, a, z)}{E\left[\operatorname{OR}(a, y) \mid r=1, a, z\right]}, \tag{8}
	\end{equation*}
	\vskip.1in
	\begin{equation*} 
		\begin{gathered}
			E\left[\widetilde{\operatorname{OR}}(a, y) \mid r=1, a, z\right]=\dfrac{f(z \mid r=0, a)}{f(z \mid r=1, a)},\\[2mm] 
			\text{where} \quad \widetilde{\operatorname{OR}}(a, y)=\dfrac{\operatorname{OR}(a, y)}{E\left[\operatorname{OR}(a, y) \mid r=1, a\right]},  \tag{9}
		\end{gathered}
	\end{equation*}
	\vskip.1in
	\begin{equation*} 
		\operatorname{OR}(a, y)=\dfrac{\widetilde{\operatorname{OR}}(a, y)}{\widetilde{\operatorname{OR}}(a, y=1)}. \tag{10} 
	\end{equation*}
	
	\paragraph*{Proof of equation (\ref{equation 7})}
	\addcontentsline{toc}{paragraph}{Proof of equation (\ref{equation 7})} 
	From Assumption \ref{shadow variable} and equation (\ref{equation 4}), we have that
	\begin{equation*}
		\begin{aligned}
			\text{right side of the equation (\ref{equation 7})}& =f(r=1 \mid a, y)  \\[3mm]
			& =f(r=1 \mid a, y, z)\\[3mm]
			& =\dfrac{f(r=1 \mid a, y=1)}{f(r=1 \mid a, y=1)+ \operatorname{OR}(a, y) f(r=0 \mid a, y=1)} \\[3mm]
			& =\dfrac{f(r=1 \mid a, y=1)}{f(r=1 \mid a, y=1)+\dfrac{f(r=0 \mid a, y) f(r=1 \mid a, y=1)}{f(r=1 \mid a, y) f(r=0 \mid a, y=1)} f(r=0 \mid a, y=1)} \\[3mm]
			& =\dfrac{f(r=1 \mid a, y=1)}{f(r=1 \mid a, y=1)+\dfrac{f(r=0 \mid a, y) f(r=1 \mid a, y=1)}{f(r=1 \mid a, y)}}\\[3mm]
			& =\dfrac{1}{1+\dfrac{f(r=0 \mid a, y)}{f(r=1 \mid a, y)}} \\[3mm]
			& =\dfrac{f(r=1 \mid a, y)}{f(r=1 \mid a, y)+f(r=0 \mid a, y)}\\[3mm]
			& =f(r=1 \mid a, y) = \text{left side of the equation (\ref{equation 7})}
		\end{aligned}
	\end{equation*}
	
	\paragraph*{Proof of equation (\ref{equation 8})} 
	\addcontentsline{toc}{paragraph}{Proof of equation (\ref{equation 8})} 
	From Assumption \ref{shadow variable} and equation (\ref{equation 4}), by the property of conditional mathematical expectation, we have that
	\begin{equation*}
		\begin{aligned}
			& E\left[\operatorname{OR}(a, y) \mid r=1, a, z\right] \\[3mm]
			&= \int \dfrac{f(r=0 \mid a, y) f(r=1 \mid a, y=1)}{f(r=1 \mid a, y) f(r=0 \mid a, y=1)} f(y \mid r=1, a, z) d \mathcal{P}(y)\\[3mm]
			&=\dfrac{f(r=1 \mid a, y=1)}{f(r=0 \mid a, y=1)} \int \dfrac{f(r=0 \mid a, y)}{f(r=1 \mid a, y)} f(y \mid r=1, a, z) d \mathcal{P}(y) \\[3mm]
			& =\dfrac{f(r=1 \mid a, y=1)}{f(r=0 \mid a, y=1)} \int \dfrac{f(r=0 \mid a, y, z)}{f(r=1 \mid a, y, z)} f(y \mid r=1, a, z) d \mathcal{P}(y) \\[3mm]
			& =\dfrac{f(r=1 \mid a, y=1)}{f(r=0 \mid a, y=1)} \int \dfrac{\dfrac{f(r=0, a, y, z)}{f(a, y, z)}}{\dfrac{f(r=1, a, y, z)}{f(a, y, z)}} \dfrac{f(y, r=1, a, z)}{f(r=1, a, z)} d \mathcal{P}(y)\\[3mm]
			& =\dfrac{f(r=1 \mid a, y=1)}{f(r=0 \mid a, y=1)} \int \dfrac{f(r=0, a, y, z)}{f(r=1, a, y, z)} \dfrac{f(y, r=1, a, z)}{f(r=1, a, z)} d \mathcal{P}(y) \\[3mm]
			& =\dfrac{f(r=1 \mid a, y=1)}{f(r=0 \mid a, y=1)} \dfrac{1}{f(r=1, a, z)} \int f(r=0, a, y, z) d \mathcal{P}(y) \\[3mm]
			& =\dfrac{f(r=1 \mid a, y=1)}{f(r=0 \mid a, y=1)} \dfrac{f(r=0, a, z)}{f(r=1, a, z)}
		\end{aligned}
	\end{equation*}
	\begin{align*}
		\text{right side of the equation (\ref{equation 8})} &= \dfrac{\operatorname{OR}(a, y) f(y \mid r=1, a, z)}{E\left[\operatorname{OR}(a, y) \mid r=1, a, z\right]} \\[3mm]
		&=\dfrac{\dfrac{f(r=0 \mid a, y) f(r=1 \mid a, y=1)}{f(r=1 \mid a, y)f(r=0 \mid a, y=1)} \quad f(y \mid r=1, a, z)}{\dfrac{f(r=1 \mid x, y=1)}{f(r=0 \mid a, y=1)} \dfrac{f(r=0, a, z)}{f(r=1, a, z)}} \\[3mm]
		&= \dfrac{\dfrac{f(r=0 \mid a, y)}{f(r=1 \mid a, y)} f(y \mid r=1, a, z)}{\dfrac{f(r=0, a, z)}{f(r=1, a, z)}}\\[3mm]
		& =\dfrac{\dfrac{f(r=0 \mid a, y, z)}{f(r=1 \mid a, y, z)} \dfrac{f(y, r=1, a, z)}{f(r=1, x, z)}}{\dfrac{f(r=0, a, z)}{f(r=1, a, z)}} \\[3mm]
		& =\dfrac{\dfrac{f(r=0, a, y, z)}{f(r=1, a, y, z)} \dfrac{f(y, r=1, a, z)}{f(r=1, a, z)}}{\dfrac{f(r=0, a, z)}{f(r=1, a, z)}} \\[3mm]
		& =\dfrac{f(r=0, a, y, z)}{f(r=0, a, z)} \\[3mm]
		& = f(y \mid r=0, a, z) = \text{left side of the equation (\ref{equation 8})}
	\end{align*}
	
	\paragraph*{Proof of equation (\ref{equation 9})}
	\addcontentsline{toc}{paragraph}{Proof of equation (\ref{equation 9})} 
	From equation (\ref{equation 6}) and equation (\ref{equation 8}) , we have that
	\begin{equation*}
		\begin{aligned}
			E\left[\operatorname{OR}(a, y) \mid r=1, a\right] &=\dfrac{f(r=1 \mid a, y=1)}{f(r=0 \mid a, y=1)} \frac{f(r=0, a)}{f(r=1, a)} \\[3mm]
			E\left[\operatorname{OR}(a, y) \mid r=1, a, z\right] & =\dfrac{f(r=1 \mid a, y=1)}{f(r=0 \mid a, y=1)} \dfrac{f(r=0, a, z)}{f(r=1, a, z)}
		\end{aligned}
	\end{equation*}
	\vskip.1in
	\begin{equation*}
		\begin{aligned}
			\text{where} \quad \widetilde{\operatorname{OR}}(a, y) &=\dfrac{\operatorname{OR}(a, y)}{E\left[\operatorname{OR}(a, y) \mid r=1, a\right]}\\[3mm]
			&=\dfrac{\dfrac{f(R=0 \mid a, y)  f(r=1 \mid a, y=1)}{f(r=1 \mid a, y) f(r=0 \mid a, y=1)}}{\dfrac{f(r=1 \mid a, y=1)  f(r=0, a)}{f(r=0 \mid a, y=1) f(r=1, a)}}\\[3mm]
			& =\dfrac{\dfrac{f(r=0 \mid a, y)}{f(r=1 \mid a, y)}}{\dfrac{f(r=0, a)}{f(r=1, a)}} \\[3mm]
			& =\dfrac{f(r=0 \mid a, y)  f(r=1, a)}{f(r=1 \mid a, y)  f(r=0, a)}\\[3mm]
			&=\dfrac{f(r=0 \mid a, y) f(r=1 \mid a)}{f(r=1 \mid a, y) f(r=0 \mid a)}
		\end{aligned}
	\end{equation*}
	\vskip.1in
	\begin{align*}
		& E\left[\widetilde{\operatorname{OR}}(a, y) \mid r=1, a, z\right] \\[3mm]
		&=\int \dfrac{f(r=0 \mid a, y) f(r=1 \mid a)}{f(r=1 \mid a, y) f(r=0 \mid a)} f(y \mid r=1, a, z) d \mathcal{P}(y)\\[3mm]
		&=\int \dfrac{f(r=0 \mid a, y, z) f(r=1, a)}{f(r=1 \mid a, y, z) f(r=0, a)} \dfrac{f(y, r=1, a, z)}{f(r=1, a, z)} d \mathcal{P}(y)\\[3mm]
		&=	\int \dfrac{f(r=0, a, y, z) f(r=1, a)}{f(r=1, a, y, z) f(r=0, a)} \dfrac{f(y, r=1, a, z)}{f(r=1, a, z)} d \mathcal{P}(y)\\[3mm]
		& =\int \dfrac{f(r=0, a, y, z) f(r=1, a)}{f(r=0, a) f(r=1, a, z)} d \mathcal{P}(y) \\[3mm]
		& =\dfrac{f(r=1, a)}{f(r=0, a) f(r=1, a, z)} \int f(r=0, a, y, z) d \mathcal{P}(y)\\[3mm]
		& =\dfrac{f(r=0, a, z) f(r=1, a)}{f(r=0, a) f(r=1, a, z)} \\[3mm]
		& =\dfrac{\dfrac{f(r=0, a, z)}{f(r=0, a)}}{\dfrac{f(r=1, a, z)}{f(r=1, a)}}\\[3mm]
		&=\dfrac{f(z \mid r=0, a)}{f(z \mid r=1, a)}
	\end{align*}

	\paragraph*{Proof of equation (\ref{equation 10})}
	\addcontentsline{toc}{paragraph}{Proof of equation (\ref{equation 10})} 
	From equation (\ref{equation 9}), we have that
	\begin{equation*}
		\begin{aligned}
			\dfrac{\widetilde{\operatorname{OR}}(a, y)}{\widetilde{\operatorname{OR}}(a, y=1)} 
			& =\dfrac{\dfrac{f(r=0 \mid a, y) f(r=1 \mid a)}{f(r=1 \mid a, y)f(r=0 \mid a)}}{\dfrac{f(r=0 \mid a, y=1) f(r=1 \mid a)}{f(r=1 \mid a, y=1)f(r=0 \mid a)}} \\[3mm]
			& =\dfrac{f(r=0 \mid a, y) f(r=1 \mid a, y=1)}{f(r=1 \mid a, y) f(r=0 \mid a, y=1)}\\[3mm]
			& ={\operatorname{OR}}(a, y)
		\end{aligned}
	\end{equation*}

	\newpage
	\subsection*{\bf Appendix B: Proofs of the Theorems}
	\addcontentsline{toc}{subsection}{Appendix B: Proofs of the theorems} 
	\subsubsection*{Proof of the Theorem \ref{theorem 1}}
	\addcontentsline{toc}{subsubsection}{Proof of the Theorem \ref{theorem 1}} 
	From Proposition \ref{proposition 2}, we have that 
	\begin{align*} \label{A.1}
		\tag{A.1}
		E\left[\widetilde{\operatorname{OR}}(a, y) \mid r=1, a, z\right]=\frac{f(z \mid r=0, a)}{f(z \mid r=1, a)}\\[3mm]			
		\widetilde{\operatorname{OR}}(a, y)=\dfrac{\operatorname{OR}(a, y)}{E\left[\operatorname{OR}(a, y) \mid r=1, a\right]}
	\end{align*}
	Under shadow variable Assumption \ref{shadow variable}, we can identify the odds ratio function $\operatorname{OR}(a, y)$. Because  $f(y \mid r=1, a, z)$ and $f(z \mid r=1, a)$ are identifiable from the observed data, for any candidate of  $\operatorname{OR}(a, y)$,  we can just need the observed data to compute the integral equation (\ref{A.1}). 
	\vskip.1in
	Suppose that $\widetilde{\operatorname{OR}}^{(1)}(a, y)$ and $\widetilde{\operatorname{OR}}^{(2)}(a, y)$ are two solutions to equation (\ref{equation 9}):
	\begin{equation*} 
		E\left[\widetilde{\operatorname{OR}}^{(k)}(a, y) \mid r=1, a, z\right]=\frac{f(z \mid r=0, a)}{f(z \mid r=1, a)} \qquad (k=1, 2),  
	\end{equation*}
	We have that
	\begin{equation*} 
		E\left[\widetilde{\operatorname{OR}}^{(1)}(a, y)-\widetilde{\operatorname{OR}}^{(2)}(a, y) \mid r=1, a, z\right]=0.
	\end{equation*}
	Condition \ref{condition 1} implies that $\widetilde{\operatorname{OR}}^{(1)}(a, y) -\widetilde{\operatorname{OR}}^{(2)}(a, y)=0$ almost surely, $\widetilde{\operatorname{OR}}^{(1)}(a, y)=\widetilde{\operatorname{OR}}^{(2)}(a, y)$ almost surely. Therefore, equation  (\ref{A.1}) has a unique solution $\operatorname{OR}(a, y)$, which means $\widetilde{\operatorname{OR}}(a, y)$ is identifiable. Based on equation (\ref{equation 10}) which is the definition of $\widetilde{\operatorname{OR}}(a, y)$, finally we can identify the odds ratio function $\operatorname{OR}(a, y)$ through as  $\operatorname{OR}(a, y)={\widetilde{\operatorname{OR}}(a, y)} / {\widetilde{\operatorname{OR}}(a, y=1)}$.
	
	\subsubsection*{Proof of the Theorem \ref{theorem 2}}
	\addcontentsline{toc}{subsubsection}{Proof of the Theorem \ref{theorem 2}} 
	
	We demonstrate some notations at the first. For simplicity, we  let $\boldsymbol{s} = \left(y, r, a,z\right)^{\mathrm{T}}$. $\boldsymbol{s}$ is a random vector denoting all the random variables. We let $g_m\left(\boldsymbol{s},\boldsymbol{\theta}\right) = g_m\left(y, r, a, z, \boldsymbol{\theta}\right)$, $\boldsymbol{\theta} \in \Theta$, $m=1,\cdots,q$  in (\ref{equation 13}).  Let
	\begin{equation*}
		\begin{aligned}
			& G( \boldsymbol{s},\boldsymbol{\theta})=\left(g_1(\boldsymbol{s},\boldsymbol{\theta}),g_2(\boldsymbol{s},\boldsymbol{\theta}) \cdots, g_q(\boldsymbol{s},\boldsymbol{\theta})\right)^{\mathrm{T}}, \quad  \boldsymbol{\theta} \in \Theta,  \\[2mm] 
			& G_0(\boldsymbol{\theta})=\left(E\left[g_1(\boldsymbol{s},\boldsymbol{\theta})\right], E\left[g_2(\boldsymbol{s},\boldsymbol{\theta})\right], \cdots, E\left[g_q(\boldsymbol{s},\boldsymbol{\theta})\right]\right)^{\mathrm{T}}, \quad  \boldsymbol{\theta} \in \Theta,  \\[2mm]
			&G_n(\boldsymbol{s}_i,\boldsymbol{\theta})=\left(g_1\left(\boldsymbol{s}_i,\boldsymbol{\theta}\right), g_2\left(\boldsymbol{s}_i,\boldsymbol{\theta}\right), \cdots, g_q\left(\boldsymbol{s}_i,\boldsymbol{\theta}\right)\right)^{\mathrm{T}}, \quad i=1,\cdots,n, \quad  \boldsymbol{\theta} \in \Theta, \\[2mm]
			& \widehat{G}_n(\boldsymbol{\theta})=\left(\frac{1}{n} \sum_{i=1}^n g_1\left(\boldsymbol{s}_i,\boldsymbol{\theta}\right), \frac{1}{n} \sum_{i=1}^n g_2\left(\boldsymbol{s}_i,\boldsymbol{\theta}\right), \cdots, \frac{1}{n} \sum_{i=1}^n g_q\left(\boldsymbol{s}_i,\boldsymbol{\theta}\right)\right)^{\mathrm{T}}, \quad  \boldsymbol{\theta} \in \Theta,
		\end{aligned}
	\end{equation*}
	\begin{equation*}
		\begin{aligned}
			& Q_0(\boldsymbol{\theta})=\left[G_0(\boldsymbol{\theta})\right]^{\mathrm{T}} W\left[G_0(\boldsymbol{\theta})\right], \\[2mm]
			& Q_n(\boldsymbol{\theta})=[\widehat{G}_n(\boldsymbol{\theta})]^{\mathrm{T}} W\left[\widehat{G}_n(\boldsymbol{\theta})\right], \\[2mm]
			&\widehat{Q}_n(\boldsymbol{\theta})=\left[\widehat{G}_n(\boldsymbol{\theta})\right]^{\mathrm{T}} \widehat{W}\left[\widehat{G}_n(\boldsymbol{\theta})\right].
		\end{aligned}
	\end{equation*}
	
	where $W$ is a positive semi-definite and symmetric $q \times q$ matrix of weights, $\widehat{W}= W(\tilde{\boldsymbol{\theta}})$. $\boldsymbol{\theta}$ is a $p$-dimensional parameter.  $\tilde{\boldsymbol{\theta}}$ is an estimator of $\boldsymbol{\theta}$ from the first step of two-step GMM, which means $\tilde{\boldsymbol{\theta}} = \mathop{\arg\min}\limits_{\boldsymbol{\theta} \in \Theta} Q_n(\boldsymbol{\theta})$.   $\boldsymbol{\theta}_0$ is the true value of the parameter $\boldsymbol{\theta}$. Let a matrix $B=\left[b_{j k}\right]$, let $\|B\|=\left(\sum_{j, k} b_{j k}^2\right)^{1/2}$ be the Euclidean norm.  Let $\nabla_{\boldsymbol{\theta}}(\cdot)$ and $\nabla_{{\boldsymbol{\theta}} {\boldsymbol{\theta}}}(\cdot)$ denote the first- and second-order derivatives with respect to $\boldsymbol{\theta}$. Let $\widehat{\boldsymbol{\theta}}=\mathop{\arg\min}\limits_{\boldsymbol{\theta} \in \Theta} \widehat{Q}_n(\boldsymbol{\theta})$.
	
	Uniform convergence in probability: $\widehat{Q}_n(\theta)$ converges uniformly in probability to $Q_0(\theta)$  means: as $n \rightarrow \infty$, 
	\begin{equation*}
		\sup _{\boldsymbol{\theta} \in \Theta}\left|\widehat{Q}_n(\boldsymbol{\theta})-Q_0(\boldsymbol{\theta})\right| \stackrel{P}{\longrightarrow} 0.
	\end{equation*}
	
	To prove the Theorem \ref{theorem 2}, we need some lemmas to prove. More details can be found in \cite{newey1994large}. 
	\begin{lemma} (GMM identification) \label{lemma 1}
		If $W$ is positive semi-definite matrix, for $G_0(\boldsymbol{\theta}_0)= \boldsymbol{0} $ and $W G_0(\boldsymbol{\theta}_0) \neq \boldsymbol{0}$ for $\boldsymbol{\theta} \neq \boldsymbol{\theta}_0$ then $Q_0(\boldsymbol{\theta})= \left[G_0(\boldsymbol{\theta})\right]^{\mathrm{T}} W\left[G_0(\boldsymbol{\theta})\right]$ has a unique minimum at $\boldsymbol{\theta}_0$.
	\end{lemma}
	\noindent {\it Proof of Lemma \ref{lemma 1}.}  Let $D$ be such that $D^{\mathrm{T}}D = W$. If $\boldsymbol{\theta} \neq \boldsymbol{\theta}_0$, then $ \boldsymbol{0} \neq W G_0(\boldsymbol{\theta}) = D^{\mathrm{T}}D G_0(\boldsymbol{\theta})$ implies $D G_0(\boldsymbol{\theta}) \neq \boldsymbol{0}$ and hence for $\boldsymbol{\theta} \neq \boldsymbol{\theta}_0$, $Q_0(\boldsymbol{\theta}) = \left[D G_0(\boldsymbol{\theta})\right]^{\mathrm{T}}\left[D G_0(\boldsymbol{\theta})\right] > Q_0(\boldsymbol{\theta}_0)$.

	\begin{lemma} \label{lemma 2}
		If the data are i.i.d., $\Theta$ is compact, $\boldsymbol{e}\left(\boldsymbol{s},\boldsymbol{\theta}\right)$ is continuous at each $\boldsymbol{\theta} \in \Theta$ with probability one, and there is $b(\boldsymbol{s})$ with $\|\boldsymbol{e}\left(\boldsymbol{s},\boldsymbol{\theta}\right)\| \leqslant b(\boldsymbol{s})$ for all $\boldsymbol{\theta} \in \Theta$ and $E[b(\boldsymbol{s})]<\infty$, then $E[\boldsymbol{e}\left(\boldsymbol{s},\boldsymbol{\theta}\right)]$ is continuous and $\sup\limits_{\boldsymbol{\theta} \in \Theta}\left\|\dfrac{1}{n} \sum\limits_{i=1}^{n} \boldsymbol{e}\left(\boldsymbol{s}_i,\boldsymbol{\theta}\right)-E[\boldsymbol{e}\left(\boldsymbol{s},\boldsymbol{\theta}\right)]\right\| \stackrel{P}{\longrightarrow} 0$.
	\end{lemma}
	\noindent {\it Proof of Lemma \ref{lemma 2}.} It is implied by Lemma 1 of \cite{tauchen1985diagnostic}.
	
	\begin{lemma} \label{lemma 3}
		If there is a function $Q_0(\boldsymbol{\theta})$ such that (i) $Q_0(\boldsymbol{\theta})$ is uniquely minimized at $\boldsymbol{\theta}_0$; (ii) $\Theta$ is compact; (iii) $Q_0(\boldsymbol{\theta})$ is continuous; (iv) $\widehat{Q}_n(\boldsymbol{\theta})$ converges uniformly in probability to $Q_0(\boldsymbol{\theta})$, then $\widehat{\boldsymbol{\theta}} \stackrel{P}{\rightarrow} \boldsymbol{\theta}_0$.
	\end{lemma}
	\noindent {\it Proof of lemma \ref{lemma 3}.}  By the definition of minimum value and the hypothesis of the Lemma \ref{lemma 3}, for any $\varepsilon>0$ we have with probability approaching one
	$$
	\widehat{Q}_n(\widehat{\boldsymbol{\theta}})<\widehat{Q}_n\left(\boldsymbol{\theta}_0\right)+\varepsilon / 3, \quad Q_0(\widehat{\boldsymbol{\theta}})<\widehat{Q}_n(\widehat{\boldsymbol{\theta}})+\varepsilon / 3, \quad \widehat{Q}_n\left(\boldsymbol{\theta}_0\right)<Q_0\left(\boldsymbol{\theta}_0\right)+\varepsilon /3.
	$$
	Therefore, with probability approaching one
	$$
	Q_0(\widehat{\boldsymbol{\theta}})<\widehat{Q}_n(\widehat{\boldsymbol{\theta}})+\varepsilon / 3<\widehat{Q}_n\left(\boldsymbol{\theta}_0\right)+2 \varepsilon / 3<Q_0\left(\boldsymbol{\theta}_0\right)+\varepsilon,
	$$
	Thus, for any $\varepsilon>0$, with probability approaching one
	$$
	Q_0(\widehat{\boldsymbol{\theta}})<Q_0\left(\boldsymbol{\theta}_0\right)+\varepsilon,
	$$
	Let $\mathscr{N}$ be any open subset of $\Theta$ containing $\boldsymbol{\theta}_0$. By $\Theta \cap \mathcal{N}^c$ compact, (i), and (iii), 
	$$
	Q_0\left(\boldsymbol{\theta}^*\right) \equiv \inf _{\boldsymbol{\theta} \in \Theta \cap \mathscr{N}^c} Q_0(\boldsymbol{\theta})>Q_0\left(\boldsymbol{\theta}_0\right),
	$$ 
	for some $ \boldsymbol{\theta}^* \in \Theta \cap \mathscr{N}^c $. 
	Thus, choosing $\varepsilon=Q_0\left(\boldsymbol{\theta}^*\right)-Q_0\left(\boldsymbol{\theta}_0\right)$, it follows that with probability approaching one
	$$
	Q_0(\widehat{\boldsymbol{\theta}})<Q_0\left(\boldsymbol{\theta}^*\right),
	$$
	hence that $\widehat{\boldsymbol{\theta}} \in \mathscr{N}$  with probability approaching one .
	By the arbitrariness of $\mathscr{N}$, 
	$$
	\widehat{\boldsymbol{\theta}} \stackrel{P}{\longrightarrow} \boldsymbol{\theta}_0.
	$$

	\vskip.1in
	\noindent {\it Proof of Theorem \ref{theorem 2}.} 
	
	Proceed by verifying the hypotheses of Lemma \ref{lemma 3}. Condition (i) in Lemma \ref{lemma 3} follows by Theorem \ref{theorem 2}  and Lemma \ref{lemma 1} . Condition (ii) in Lemma \ref{lemma 3} holds by Theorem \ref{theorem 2}. By Lemma \ref{lemma 2} applied to $\boldsymbol{e}\left(\boldsymbol{s},\boldsymbol{\theta}\right) = G(\boldsymbol{s},\boldsymbol{\theta}) $,  one has $\sup\limits_{\boldsymbol{\theta} \in \Theta}\left\|\widehat{G}_n(\boldsymbol{\theta})-G_0(\boldsymbol{\theta})\right\| \stackrel{p}{\rightarrow} 0$ and $G_0(\boldsymbol{\theta})$ is continuous. Thus, (iii) in Lemma \ref{lemma 3} holds by $Q_0(\boldsymbol{\theta})=\left[G_0(\boldsymbol{\theta})\right]^{\mathrm{T}} W\left[G_0(\boldsymbol{\theta})\right]$ continuous. By $\Theta$ compact, $G_0(\boldsymbol{\theta})$ is bounded on $\Theta$. 
	
	Suppose that $\widehat{W}$ converges to $W$ in probability. By triangle and Cauchy-Schwartz inequalities,
	
	\begin{equation*}
		\begin{aligned}
			&\left|\widehat{Q}_n(\boldsymbol{\theta})-Q_0(\boldsymbol{\theta})\right| \\
			& \leqslant\left|\left[\widehat{G}_n(\boldsymbol{\theta})-G_0(\boldsymbol{\theta})\right]^{\mathrm{T}} \widehat{W}\left[\widehat{G}_n(\boldsymbol{\theta})-G_0(\boldsymbol{\theta})\right]\right|+ \left|\left[G_0(\boldsymbol{\theta})\right]^{\mathrm{T}} \left(\widehat{W}+\widehat{W}^{\mathrm{T}}\right)\left[\widehat{G}_n(\boldsymbol{\theta})-G_0(\boldsymbol{\theta})\right]\right| \\
			&+\left|[G_0(\boldsymbol{\theta})]^{\mathrm{T}}(\widehat{W}-W) G_0(\boldsymbol{\theta})\right| \\
			& \leqslant\|\widehat{G}_n(\boldsymbol{\theta})-G_0(\boldsymbol{\theta})\|^2\|\widehat{W}\|+2\left\|G_0(\boldsymbol{\theta})\right\|\|\widehat{G}_n(\boldsymbol{\theta})-G_0(\boldsymbol{\theta})\|\|\widehat{W}\| \\
			& \quad+\left\|G_0(\boldsymbol{\theta})\right\|^2\|\widehat{W}-W\| ,
		\end{aligned}
	\end{equation*}
	so that 
	\begin{equation*}
		\sup\limits_{\boldsymbol{\theta} \in \Theta}\left|\widehat{Q}_n(\boldsymbol{\theta})-Q_0(\boldsymbol{\theta})\right| \stackrel{P}{\longrightarrow} 0 \text {, and (iv) in Lemma \ref{lemma 3} holds. }
	\end{equation*}
	With $W$ being the $I_{q \times q}$ identity matrix, we can prove that
	$$
	\tilde{\boldsymbol{\theta}} \stackrel{P}{\longrightarrow} \boldsymbol{\theta}_0.
	$$
	which implies $\widehat{W} \stackrel{P}{\longrightarrow} W$.

	\subsubsection*{Proof of the Theorem \ref{theorem 3}}
	\addcontentsline{toc}{subsubsection}{Proof of the Theorem \ref{theorem 3}} 
	Under the assumptions in Theorem \ref{theorem 2}, $\nabla_{\boldsymbol{\theta}} \widehat{Q}_n(\widehat{\boldsymbol{\theta}})=0$ with probability approaching 1. By Taylor's expansion around $\boldsymbol{\theta}_0$, 
	$$
	\nabla_{\boldsymbol{\theta}} \widehat{Q}_n(\widehat{\boldsymbol{\theta}}) - \nabla_{\boldsymbol{\theta}} \widehat{Q}_n(\boldsymbol{\theta}_0)= \nabla_{{\boldsymbol{\theta}} {\boldsymbol{\theta}}} \widehat{Q}_n(\boldsymbol{\theta}^*) (\widehat{\boldsymbol{\theta}}- \boldsymbol{\theta}_0), 
	$$
	where $\boldsymbol{\theta}^*$ is between $\widehat{\boldsymbol{\theta}}$ and $\boldsymbol{\theta}_0$. Let $\widehat{H}(\widehat{\boldsymbol{\theta}}) = \nabla_{\boldsymbol{\theta}} \widehat{G}_n(\widehat{\boldsymbol{\theta}})$. Multiplying through by $\sqrt{n}$,  and solving gives
	$$
	\sqrt{n}\left(\widehat{\boldsymbol{\theta}}-\boldsymbol{\theta}_0\right)=-\left\{\left[\widehat{H}(\widehat{\boldsymbol{\theta}})\right]^{\mathrm{T}} \widehat{W} \left[\widehat{H}(\boldsymbol{\theta}^*)\right]\right\}^{-1}\left[\widehat{H}(\widehat{\boldsymbol{\theta}})\right]^{\mathrm{T}} \widehat{W} \left[\sqrt{n} \widehat{G}_n(\boldsymbol{\theta}_0) \right] ,
	$$
	By (iii) in Condition \ref{condition 3}, $\widehat{H}(\widehat{\boldsymbol{\theta}}) \stackrel{P}{\longrightarrow} H$ and $\widehat{H}(\boldsymbol{\theta}^*) \stackrel{P}{\longrightarrow} H$, so that by (iv) and continuity of matrix inversion, 
	$$
	- \left\{\left[\widehat{H}(\widehat{\boldsymbol{\theta}})\right]^{\mathrm{T}} \widehat{W} \left[\widehat{H}(\boldsymbol{\theta}^*)\right]\right\}^{-1} \left[\widehat{H}(\widehat{\boldsymbol{\theta}})\right]^{\mathrm{T}} \widehat{W} \stackrel{P}{\longrightarrow}  \left(H^{\mathrm{T}} W H\right)^{-1}H^{\mathrm{T}} W ,
	$$
	Based on the Central Limit Theorem,
	$$
	\left[\sqrt{n} \widehat{G}_n(\boldsymbol{\theta}_0) \right]  \stackrel{\mathscr{L}}{\longrightarrow} N(\boldsymbol{0},\Omega),
	$$
	the conclusion then follows by the Slutzky theorem.
	$$
	\sqrt{n}(\widehat{\boldsymbol{\theta}}-\boldsymbol{\theta}_0) \stackrel{\mathscr{L}}{\longrightarrow} N\left(\boldsymbol{0}, \left(H^{\mathrm{T}} W H\right)^{-1} H^{\mathrm{T}} W \Omega W H \left(H^{\mathrm{T}} W H\right)^{-1}\right).
	$$
	
	\subsubsection*{Proof of the Theorem \ref{theorem 4}}
	\addcontentsline{toc}{subsubsection}{Proof of the Theorem \ref{theorem 4}} 
	Under the assumptions of Theorem \ref{theorem 2} and Theorem \ref{theorem 3}, we have the following proof.
	
	By consistency of $\widehat{\boldsymbol{\theta}}$ there is $\lambda_n \rightarrow 0$ such that $\|\widehat{\boldsymbol{\theta}}-\boldsymbol{\theta}_0\| \leqslant \lambda_n$ with probability approaching one. 
	
	Let $\Lambda_n(\boldsymbol{s})=\sup_{\|\widehat{\boldsymbol{\theta}}-\boldsymbol{\theta}_0\| \leqslant \lambda_n} \|G(\boldsymbol{s},\boldsymbol{\theta}) G(\boldsymbol{s},\boldsymbol{\theta})^{\mathrm{T}}-G(\boldsymbol{s},\boldsymbol{\theta}_0)G(\boldsymbol{s},\boldsymbol{\theta}_0)^{\mathrm{T}}\|$.
	
	By continuity of $G(\boldsymbol{s},\boldsymbol{\theta}) G(\boldsymbol{s},\boldsymbol{\theta})^{\mathrm{T}}$ at $\boldsymbol{\theta}_0$, $\Lambda_n(\boldsymbol{s}) \rightarrow 0$ with probability one, while by the dominance condition, for $n$ large enough $\Lambda_n(\boldsymbol{s}) \leqslant 2 \sup_{\boldsymbol{\theta} \in \mathcal{N}}\|G(\boldsymbol{s},\boldsymbol{\theta}) G(\boldsymbol{s},\boldsymbol{\theta})^{\mathrm{T}}\|$. 
	
	Then by the dominated convergence theorem, $E\left[\Lambda_n(\boldsymbol{s})\right] \rightarrow 0$ , so by the Markov's inequality, 
	$$
	P\left(\left| \frac{1}{n} \sum_{i=1}^n \Lambda_n(\boldsymbol{s}_i)\right|>\varepsilon\right) \leqslant \frac{E\left[\Lambda_n(\boldsymbol{s})\right]}{\varepsilon} \rightarrow 0 ,
	$$
	for all $\varepsilon>0$, giving $\frac{1}{n} \sum\limits_{i=1}^n \Lambda_n(\boldsymbol{s}_i) \stackrel{P}{\longrightarrow} 0$. 
	By Khinchin's law of large numbers, 
	$$
	\frac{1}{n} \sum_{i=1}^n G(\boldsymbol{s}_i,\boldsymbol{\theta}_0)G(\boldsymbol{s}_i,\boldsymbol{\theta}_0)^{\mathrm{T}} \stackrel{P}{\longrightarrow} E\left[G(\boldsymbol{s},\boldsymbol{\theta}_0)G(\boldsymbol{s},\boldsymbol{\theta}_0)^{\mathrm{T}}\right] ,
	$$ 
	Also, with probability approaching one,
	\begin{equation*}
		\begin{aligned}
			\left\| \frac{1}{n} \sum_{i=1}^n G(\boldsymbol{s}_i,\widehat{\boldsymbol{\theta}})G(\boldsymbol{s}_i,\widehat{\boldsymbol{\theta}})^{\mathrm{T}} - \frac{1}{n} \sum_{i=1}^n G(\boldsymbol{s}_i,\boldsymbol{\theta}_0)G(\boldsymbol{s}_i,\boldsymbol{\theta}_0)^{\mathrm{T}}  \right\|
			& \leqslant \frac{1}{n} \sum_{i=1}^n\left\| 
			G(\boldsymbol{s}_i,\widehat{\boldsymbol{\theta}})G(\boldsymbol{s}_i,\widehat{\boldsymbol{\theta}})^{\mathrm{T}}-G(\boldsymbol{s}_i,\boldsymbol{\theta}_0) G(\boldsymbol{s}_i,\boldsymbol{\theta}_0)^{\mathrm{T}} \right\| \\
			&\leqslant \frac{1}{n} \sum_{i=1}^n \Lambda_n(\boldsymbol{s}_i) \stackrel{P}{\longrightarrow} 0 .
		\end{aligned}
	\end{equation*}
	So by the triangle inequality, we have that $$
	\frac{1}{n} \sum_{i=1}^n G(\boldsymbol{s}_i,\widehat{\boldsymbol{\theta}})G(\boldsymbol{s}_i,\widehat{\boldsymbol{\theta}})^{\mathrm{T}} \stackrel{P}{\longrightarrow} E\left[G(\boldsymbol{s},\boldsymbol{\theta}_0)G(\boldsymbol{s},\boldsymbol{\theta}_0)^{\mathrm{T}}\right] ,
	$$
	which means $\widehat{\Omega} \stackrel{P}{\longrightarrow} \Omega$.
	Theorem \ref{theorem 2} and Theorem \ref{theorem 3} implies that $\widehat{H} \stackrel{P}{\longrightarrow} H$, $\widehat{W} \stackrel{P}{\longrightarrow} W$. The conclusion follows by (iv) of Condition \ref{condition 3} and continuity of matrix inversion and multiplication.

\end{document}